\documentclass[12pt]{article}

\usepackage{amsbsy}
\usepackage{amsmath}
\usepackage{amssymb}
\usepackage{amsthm}
\usepackage{color}
\usepackage{enumerate}
\usepackage{enumitem}
\usepackage{geometry}
\usepackage{graphicx}
\usepackage{caption}
\usepackage{subcaption}
\usepackage{float}
\usepackage{natbib}
\usepackage[hidelinks]{hyperref}
\usepackage{setspace}
\usepackage{mathrsfs}
\usepackage{physics}
\usepackage{url}
\usepackage{comment}
\usepackage{longtable}            
\usepackage{rotating}
\usepackage[monochrome]{xcolor}

\newcommand{\blind}{1}

\theoremstyle{plain}
\newtheorem{thm}{Theorem}

\DeclareMathOperator*{\argmin}{arg\,min}

\def\bp{\mathbf{p}}
\def\bx{\mathbf{x}}
\def\bz{\mathbf{z}}
\def\bS{\mathbf{S}}
\def\bmu{\boldsymbol{\mu}}
\def\bSigma{\boldsymbol{\Sigma}}
\def\bOmega{\boldsymbol{\Omega}}

\begin{document}

\def\spacingset#1{\renewcommand{\baselinestretch}%
{#1}\small\normalsize} \spacingset{1}


\if1\blind
{
  \title{\bf Compositional Graphical Lasso Resolves the Impact of Parasitic Infection on Gut Microbial Interaction Networks in a Zebrafish Model}
  \author{Chuan Tian$^\text{a}$, Duo Jiang$^\text{a}$, Austin Hammer$^\text{b}$, \\
  Thomas Sharpton$^\text{a,b}$, and Yuan Jiang$^\text{a}$\thanks{Yuan Jiang is the corresponding author (E-mail: yuan.jiang@oregonstate.edu). This research is supported in part by the grant R01 GM126549 from the National Institutes of Health (NIH).}\\
  $^\text{a}$Department of Statistics, Oregon State University, Corvallis, OR\\
  $^\text{b}$Department of Microbiology, Oregon State University, Corvallis, OR\\
  }
  
  \maketitle
} \fi

\if0\blind
{
  \bigskip
  \bigskip
  \bigskip
  \begin{center}
    {\Large\bf Compositional Graphical Lasso Resolves the Impact of Parasitic Infection on Gut Microbial Interaction Networks in a Zebrafish Model}
\end{center}
  \medskip
} \fi

\bigskip
\begin{abstract}

Understanding how microbes interact with each other is key to revealing the underlying role that microorganisms play in the host or environment and to identifying microorganisms as an agent that can potentially alter the host or environment. For example, understanding how the microbial interactions associate with parasitic infection can help resolve potential drug or diagnostic test for parasitic infection. To unravel the microbial interactions, existing tools often rely on graphical models to infer the conditional dependence of microbial abundances to represent their interactions. However, current methods do not simultaneously account for the discreteness, compositionality, and heterogeneity inherent to microbiome data. Thus, we build a new approach called ``compositional graphical lasso'' upon existing tools by incorporating the above characteristics into the graphical model explicitly. We illustrate the advantage of compositional graphical lasso over current methods under a variety of simulation scenarios and on a benchmark study, the \textit{Tara} Oceans Project. Moreover, we present our results from the analysis of a dataset from the Zebrafish Parasite Infection Study, which aims to gain insight into how the gut microbiome and parasite burden covary during infection, thus uncovering novel putative methods of disrupting parasite success. 
{\color{blue}Our approach identifies changes in interaction degree between infected and uninfected individuals for three taxa, \textit{Photobacterium}, \textit{Gemmobacter}, and \textit{Paucibacter}, which are inversely predicted by other methods. Further investigation of these method-specific taxa interaction changes reveals their biological plausibility. In particular, we speculate on the potential pathobiotic roles of \textit{Photobacterium} and \textit{Gemmobacter} in the zebrafish gut, and the potential probiotic role of \textit{Paucibacter}. Collectively, our analyses demonstrate that compositional graphical lasso provides a powerful means of accurately resolving interactions between microbiota and can thus drive novel biological discovery.}

\end{abstract}

\noindent%
\vfill

\newpage
\spacingset{1.5} 
\section{Introduction}
\label{Sec-1}

\subsection{Background}
\label{Sec-1.1}

Microorganisms are ubiquitous in nature and manage critical ecosystem services, ranging from global nutrient cycling to human health. Microbes do not exist in a vacuum in nature, but instead are members of diverse communities known as microbiomes, wherein microbes may interact with other microbes of different species. These microbial interactions may be favorable or antagonistic and are crucial for the successful establishment and maintenance of a microbial community, and frequently result in important pathogenic or beneficial effect to the host or environment \citep{braga2016microbial}. A thorough understanding of how microbe interact with one another is critical to uncovering the underlying role microorganisms play in the host or environment. 
However, it is a highly challenging biological task as it is estimated that only 1\% of bacteria are cultivatable. The inability to culture the majority of microbial species has motivated the use of culture-independent methods for microbiome studies in different environments \citep{faust2012microbial, tang2019microbial}. 

Fortunately, recent innovations in in situ DNA sequencing provide opportunities to infer how microbes interact with one another or their environments. Modern studies on microbial interactions frequently rely on DNA sequencing techniques through the bioinformatic analysis of taxonomically diagnostic genetic markers (e.g., 16S rRNA) sequenced directly from a sample.  The counts of the taxonomically diagnostic genetic markers can be used to represent the abundance of microbial species, e.g., Operational Taxonomic Units (OTUs) or phylotypes (e.g., genera), in a sample. Here, the frequency with which a taxon’s marker is observed in a sequence library represents its relative abundance in the community. When such abundance data are available from many communities, interactions among microbiota can be inferred through \textcolor{blue}{statistical} correlation analysis \citep{faust2012microbial}. \textcolor{blue}{For example}, if the relative abundances of two microbial taxa are statistically correlated, then it is inferred that they interact on some level. This approach has been used to document interaction networks in the healthy human microbiome \citep{faust2012microbial}, as well as free-living microbial communities \citep{freilich2010large}, and has been useful for generating hypotheses of host-microbiome interaction \citep{morgan2015associations}.

\subsection{Methodological Innovations}
\label{Sec-1.2}

Despite the great potential of microbiome interaction networks as a tool to advance microbiome research, the power of this approach has been limited by the availability of effective statistical methods and computationally efficient estimation techniques. Microbial abundance data possess a few important features that pose tremendous challenges to standard statistical tools. First, the data are represented as compositional counts of the 16S rRNA sequences because the total count of sequences per sample is predetermined by how deeply the sequencing is conducted, a concept named sequencing depth. The counts only carry information about the relative abundances of the taxa instead of their absolute abundances. Second, the sequencing depth is always finite and often varies considerably across samples in a microbiome dataset. Thus, the observed relative abundance of a taxon in a sample is only an estimator of its true relative abundance with the variance depending on sample-specific sequencing depth, causing the ``heteroscedasticity'' issue \citep{mcmurdie2014waste}. Third, the data are high-dimensional in nature. It is likely that the number of taxa is far more than the number of samples in any biological experiment.


When such abundance data are available, one common strategy to resolve the interactions among microbial taxa is to use correlation-type analyses. For example, after the sample correlations are calculated between \textcolor{blue}{the relative abundances of} each pair of microbial taxa, a threshold is then applied such that an interaction is deemed present if the sample correlation exceeds the threshold. More \textcolor{blue}{recently developed methods} have started to \textcolor{blue}{account for} the compositional feature and aim to construct sparse networks for the absolute abundances instead of relative abundances, \textcolor{blue}{including} SparCC \citep{friedman2012inferring}, CCLasso \citep{fang2015cclasso}, and REBACCA \citep{ban2015investigating}, \textcolor{blue}{among others.}

\textcolor{blue}{The above-mentioned network methods based on marginal correlations} could lead to spurious correlations that are caused by confounding factors such as other \textcolor{blue}{species} in the same community. \textcolor{blue}{To eliminate the detection of spurious correlations}, interactions among taxa can be modeled through their conditional dependencies given the other taxa. The Gaussian graphical model \textcolor{blue}{is a classical approach to modeling} the conditional dependency, \textcolor{blue}{in which the conditional dependency is determined by the nonzero entries of the inverse covariance matrix of a multivariate normal distribution used to model the data.} Graphical lasso \citep{yuan2007model, banerjee2008model, friedman2008sparse} and neighborhood selection \citep{meinshausen2006high} are two commonly used methods to estimate sparse inverse covariance matrix under the \textcolor{blue}{high-dimensional} Gaussian graphical model. However, \textcolor{blue}{when applied to microbiome abundance data, its multivariate normality assumption was violated by either the count or the compositional features of the data.}

Several methods have been proposed to infer microbial conditional dependence networks based on Gaussian graphical models, such as SPIEC-EASI \citep{kurtz2015sparse}, gCoda \citep{fang2017gcoda}, CD-trace \citep{yuan2019compositional}, and SPRING \citep{yoon2019microbial}. In order to transform the discrete counts to continuous variables and to remove the compositionality constraint, all these methods take the centered log-ratio transformation \citep{aitchison1986statistical} on the observed counts as their first step. However, the centered log-ratio transformation suffers from an undefined inverse covariance matrix of the transformed data. To partially address this issue, these methods impose a sparsity assumption on the inverse covariance matrix and adding an $L_1$-norm penalty to their objective functions. Additionally, these methods ignore the heteroscedasticity issue as they simply treat the observed relative abundances as the truth. Ignoring the heteroscedasticity issue could impact downstream analysis including constructing microbial interaction networks \citep{mcmurdie2014waste}.

In this article, we provide a new statistical tool to help unleash the full potential of microbiome interaction networks as a research tool in the microbiome field. We adopt the logistic normal multinomial distribution to model the compositional count data \citep{aitchison1986statistical, billheimer2001statistical, xia2013logistic}. Compared to previous methods, this model accounts for the heteroscedasticity issue as the sequencing depth is treated as the number of trials in the multinomial distribution. Additionally, the additive log-ratio transformation applied to the multinomial probabilities results in a well-defined inverse covariance matrix in contrast to the centered log-ratio transformation. Based on this model, we develop an efficient algorithm that iterates between Newton-Raphson and graphical lasso for estimating a sparse inverse covariance matrix. We call this new approach ``compositional graphical lasso''. We establish the theoretical convergence of the algorithm and illustrate the advantage of compositional graphical lasso in comparison to current methods under a variety of simulation scenarios. We further apply the developed method to the data from the Zebrafish Parasite Infection study \citep{gaulke2019longitudinal} (see Section \ref{Sec-1.3}) to investigate how microbial interactions associate with parasite infection.


\subsection{Zebrafish Parasite Infection Study}
\label{Sec-1.3}

Helminth parasites represent a significant threat to the health of human and animal populations, and there is a growing need for tools to treat, diagnose, and prevent these infections. A growing body of evidence points to the gut microbiome as an agent that interacts with parasites to influence their success in the gut. To clarify how the gut microbiome varies in accordance with parasitic infection dynamics, the Zebrafish Parasite Infection Study was a recent effort \citep{gaulke2019longitudinal} conducted at Oregon State University, which assessed the association of an intestinal helminth of zebrafish, \textit{Pseudocapillaria tomentosa}, and the gut microbiome of 210 4-month-old 5D line zebrafish. Among these fish, 105 were exposed to \textit{P.\ tomentosa} and the remaining 105 were unexposed controls. At each of the seven time points after exposure, a randomly selected group of 30 fish (15 exposed and 15 unexposed) were euthanized and fecal samples were collected. The parasite burden and tissue damage in \textit{P.\ tomentosa}-infected fish were also monitored over 12 weeks of infection.

Previous analyses \citep{gaulke2019longitudinal} of the Zebrafish Parasite Infection Study data have revealed that parasite exposure, burden, and intestinal lesions were correlated with gut microbial diversity. They also identified individual taxa whose abundance associated with parasite burden, suggesting that gut microbiota may influence \textit{P.\ tomentosa} success. Numerous associations between taxon abundance, burden, and gut pathologic changes were also observed, indicating that the magnitude of microbiome disruption during infection varies with infection severity. However, it remains unclear how parasite success may disrupt or be modulated by the microbial interactions in the gut. Understanding how the microbial interactions associate with parasitic infection can help resolve potential drug or diagnostic test for parasitic infection.

\subsection{Benchmarking and Novel Biological Discoveries}
\label{Sec-1.4}

To evaluate the performance of our method and to benchmark it against previously available tools, we take advantage of a unique data resource provided by the \textit{Tara} Oceans Project on ocean planktons. This data set is particularly suitable for method comparison because it enjoys an experimentally validated sub-network of plankton interactome that has served as a gold standard for method benchmarking. Compared to other methods, compositional graphical lasso performs better in reconstructing the microbial interactions that are validated by the literature. In addition, it performs the best in picking up the keystone taxa, which are those with an excessive number of interactions with other taxa in the literature, such as \textit{Amoebophyra} \citep{chambouvet2008control}, \textit{Blastodinium} \citep{skovgaard2012parasitic}, \textit{Phaeocystis} \citep{verity2007current}, and \textit{Syndinium} \citep{skovgaard2005phylogenetic}. All these genera are well-described keystone organisms in marine ecosystems. Finally, compositional graphical lasso affords an opportunity to resolve novel modulators of community composition. For example, as one of a few described genera within the syndinean dinoflagellates---an enigmatic lineage with abundant diversity in marine environmental clone libraries, \textit{Euduboscquella} \citep{bachvaroff2012molecular} ranked high in the degree distribution uniquely by compositional graphical lasso. 

To investigate the role of the gut microbiome plays in the parasite infections, we apply compositional graphical lasso to the Zebrafish Parasite Infection Study data. {\color{blue}Interestingly, compositional graphical lasso identifies changes in interaction degree between infected and uninfected individuals for three taxa, \textit{Photobacterium}, \textit{Gemmobacter}, and \textit{Paucibacter}, which are inversely predicted by other methods. Further investigation of these method-specific taxa interaction changes reveals their biological plausibility, and provides insight into their relevance in the context of parasite-linked changes in the zebrafish gut microbiome. In particular, based on our observations, we speculate on the potential pathobiotic roles of \textit{Photobacterium} and \textit{Gemmobacter} in the zebrafish gut, and the potential probiotic role of \textit{Paucibacter}. Future studies should seek to experimentally validate the ecological roles of \textit{Photobacterium}, \textit{Gemmobacter}, and \textit{Paucibacter} in the zebrafish gut, including their impacts on the rest of the microbial community and their roles in infection induced tissue damage.}

\section{Compositional Graphical Lasso}
\label{Sec-2}

\subsection{Logistic Normal Multinomial Model}
\label{Sec-2.1}

Consider a microbiome abundance dataset with $n$ independent samples, each of which is composed of the observed counts of $K + 1$ taxa, denoted by $\bx_i = (x_{i,1}, \ldots, x_{i,K+1})'$ for the $i$-th sample, $i = 1,\ldots,n$. Due to the compositional property of the data, the total count of all taxa for each sample $i$ is a fixed number, denoted by $M_i$. Naturally, a multinomial distribution is imposed on the observed counts:
\begin{equation} 
\bx_i | \bp_i \sim \text{Multinomial}(M_i; p_{i,1}, \ldots, p_{i,K +1}), \label{multinomial}
\end{equation}
where $\bp_i = (p_{i,1}, \ldots, p_{i,K+1})'$ are the multinomial probabilities for all taxa and $\sum_{k=1}^{K+1} p_{i,k} = 1$.

To apply the additive log-ratio transformation \citep{aitchison1986statistical} on the multinomial probabilities, we choose one taxon, without loss of generality the $(K+1)$-th taxon, as a reference to which all the other tax are compared. The transformed multinomial probabilities are given by
\begin{equation}
z_{i,k} = \log (\frac{p_{i, k}}{p_{i, K + 1}}),\ i = 1,\ldots,n, \ k = 1,\ldots,K. \label{log.ratio.transformation}
\end{equation}
Let $\mathbf{z}_i = (z_{i,1}, \ldots, z_{i,K})'$ for $i = 1,\ldots,n$, and further assume that they follow an i.i.d. multivariate normal distribution
\begin{equation}
\bz_1, \ldots, \bz_n \stackrel{i.i.d.}{\sim} N (\boldsymbol\mu, \boldsymbol\Sigma), \label{logistic.normal}
\end{equation}
where $\boldsymbol\mu$ is the mean and $\boldsymbol\Sigma$ is the covariance matrix. Let $\boldsymbol\Omega = \boldsymbol\Sigma^{-1}$ be the inverse covariance matrix or the precision matrix.

The above model given in (\ref{multinomial})--(\ref{logistic.normal}) is often referred to as the logistic normal multinomial model. In this model, a multinomial distribution is imposed on the compositional counts, which is the distribution of the observed data given the multinomial probabilities. In addition, to capture the variation of the multinomial probabilities across samples, we impose a logistic normal distribution on the multinomial probabilities as a prior distribution. We thereby obtain as our final model the logistic normal multinomial model, which is a hierarchical model with two levels.

The logistic normal multinomial model has a long history in modeling compositional count data and it has also been applied to analyze microbiome abundance data. For example, \citet{xia2013logistic} proposed a penalized regression under this model to identify a subset of covariates that are associated with the taxon composition. Our objective is different from \citet{xia2013logistic} as we aim to reveal the microbial interaction network by finding a sparse estimator of the inverse covariance matrix $\bOmega$ in (\ref{logistic.normal}). It is also noteworthy that \cite{jiang2020microbial} has the same objective as ours. However, \cite{jiang2020microbial} did not make full use of the logistic normal multinomial model as it focused on correcting the bias of a naive estimator of the $\bSigma$ that does not require the logistic normal part of the model. By contrast, we aim to find an estimator of $\bOmega$ directly based on the logistic normal multinomial model.


\subsection{Objective Function}
\label{Sec-2.2}

{\color{blue}From the logistic normal multinomial model in (\ref{multinomial})--(\ref{logistic.normal}), we aim to derive an objective function of $\bOmega$ to estimate the microbial interaction network. To this end, we take a two-step procedure similar to SPIEC-EASI \citep{kurtz2015sparse}. In the first step, we find estimated values of $\bz_1,\ldots,\bz_n$ based on the logistic normal multinomial model given $\mu$ and $\bSigma$; in the second step, we find an estimate of $\bOmega$ based on the estimated values of $\bz_1,\ldots,\bz_n$.

In the first step, we consider the posterior distribution of $\bz_1,\ldots,\bz_n$ given the data $\bx_1,\ldots,\bx_n$ and find the maximum a posteriori (MAP) estimates of $\bz_1,\ldots,\bz_n$. For $i=1,\ldots,n$, the logarithm of the posterior density function of $\bz_i$ given $\bx_i$ is
\begin{align*} 
& \log[f_{\bmu,\bOmega}(\bz_i | \bx_i)] \propto \log[f_{\bmu,\bOmega}(\bx_i, \bz_i)] \\
\propto{}& \sum_{k=1}^{K+1} x_{i,k} \log p_{i,k} + \frac12 \log[\det(\bOmega)] - \frac12 (\bz_i - \bmu)' \bOmega (\bz_i - \bmu) \\
={} &\sum_{k=1}^{K} x_{i,k} z_{i,k} - M_i \log(\sum_{k=1}^K e^{z_{i,k}} + 1) + \frac12 \log[\det(\bOmega)] - \frac12 (\bz_i - \bmu)' \bOmega (\bz_i - \bmu),
\end{align*}
where $\propto$ denotes that two quantities are equal up to a term not depending on $\bz_i$. In the above derivation, we ignored the marginal density function of $\bx_i$ that does not depend on $\bz_i$, which does not affect the estimation of $\bz_i$.}

By independence between all the samples, the logarithm of the posterior density function of $(\bz_1,\ldots,\bz_n)$ given the data $(\bx_1,\ldots,\bx_n)$ can be written as (again, ignoring a term independent of $\bz_1,\ldots,\bz_n$):
\begin{equation} 
\sum_{i=1}^n \left[\sum_{k=1}^{K} x_{i,k} z_{i,k} - M_i \log(\sum_{k=1}^K e^{z_{i,k}} + 1)\right] + \frac{n}2 \log[\det(\bOmega)] - \frac12 \sum_{i=1}^n (\bz_i - \bmu)' \bOmega (\bz_i - \bmu). \label{objective.function.1}
\end{equation}
Given the values of the multivariate normal parameters $\bmu$ and $\bOmega$, one can maximize (\ref{objective.function.1}) with respect to $(\bz_1,\ldots,\bz_n)$. This leads to the MAP estimator $(\hat\bz_1,\ldots,\hat\bz_n)$.

{\color{blue}In the second step, we find a sparse inverse covariance estimator of $\bOmega$ based on the estimate values $(\hat\bz_1,\ldots,\hat\bz_n)$. Hereby, we use the graphical lasso estimator, which minimizes the $L_1$ penalized negative log-likelihood function as follows:
\begin{equation}
-\frac{1}2 \log[\det(\bOmega)] + \frac1{2n} \sum_{i=1}^n (\hat\bz_i - \bmu)' \bOmega (\hat\bz_i - \bmu) + \lambda \|\bOmega\|_1. \label{objective.function.2}
\end{equation}

It turns out that the above two-step procedure is equivalent to minimizing an overall objective function with respect to both $\bz_1,\ldots,\bz_n$ and $(\bmu,\bOmega)$:
\begin{align}
\ell(\bz_1,\ldots,\bz_n,\bmu,\bOmega) ={}& -\frac1n\sum_{i=1}^n \left[\sum_{k=1}^{K} x_{i,k} z_{i,k} - M_i \log(\sum_{k=1}^K e^{z_{i,k}} + 1)\right] \notag\\
& -\frac12 \log[\det(\bOmega)] + \frac{1}{2n} \sum_{i=1}^n (\bz_i - \bmu)' \bOmega (\bz_i - \bmu) + \lambda \|\bOmega\|_1. \label{objective.function}
\end{align}
In other words, $\bz_1,\ldots,\bz_n$, $\bmu$, and $\bOmega$ are all treated as unknown parameters in the minimization of the objective function (\ref{objective.function}).

It is noteworthy that similar ideas to the above two-step procedure have been used in existing microbial interaction network estimation methods. For example, SPIEC-EASI is also a two-step procedure. In its first step, the abundance counts are converted into their centered log-ratio transformed data; in its second step, either graphical lasso or neighborhood selection is applied to estimate a sparse inverse covariance matrix based on the centered log-ratio transformed data in the first step.

Although both our method and SPIEC-EASI can be regarded as two-step procedures, we underline two important distinctions between them. First, our method is based on the additive log-ratio transformation and SPIEC-EASI uses the centered log-ratio transformation. As mentioned in the Introduction, the centered log-ratio transformation suffers from an undefined inverse covariance matrix of the transformed data while the additive log-ratio transformation does not. Second, in the first step, our method accounts for the sequencing depths [$M_i$'s in (\ref{multinomial})] and further the uncertainty of the observed relative abundances, and thus addresses the heteroscedasticity issue. However, the heteroscedasticity issue is ignored in SPIEC-EASI.} 


\subsection{Computational Algorithm}
\label{Sec-2.3}

The objective function (\ref{objective.function})  naturally includes three sets of parameters $(\bz_1,\ldots,\bz_n)$, $\bmu$, and $\bOmega$, which motivates us to apply a block coordinate descent algorithm. A block coordinate descent algorithm minimizes the objective function iteratively for each set of parameters given current values of the other sets. Given the initial values $(\bz_1^{(0)},\ldots,\bz_n^{(0)})$, $\bmu^{(0)}$, and $\bOmega^{(0)}$, a block coordinate algorithm repeats the following steps cyclically for iteration $t= 0,1,2,\ldots$ until the algorithm converges.
\begin{enumerate}[nosep]
\item Given $\bmu^{(t)}$ and $\bOmega^{(t)}$, find $(\bz_1^{(t+1)},\ldots,\bz_n^{(t+1)})$ that maximizes (\ref{objective.function}).
\item Given $(\bz_1^{(t+1)},\ldots,\bz_n^{(t+1)})$ and $\bOmega^{(t)}$, find $\bmu^{(t+1)}$ that maximizes (\ref{objective.function}).
\item Given $(\bz_1^{(t+1)},\ldots,\bz_n^{(t+1)})$ and $\bmu^{(t+1)}$, find $\bOmega^{(t+1)}$ that maximizes (\ref{objective.function}).
\end{enumerate}

Below, we will present the details of this algorithm in each iteration. For the initial values $(\bz_1^{(0)},\ldots,\bz_n^{(0)})$, we use their maximum likelihood estimators from the multinomial distribution, i.e.,
\[ {z}_{i,k}^{(0)} = \log (\frac{x_{i, k}}{x_{i, K + 1}}),\ i = 1,\ldots,n, \ k = 1,\ldots,K.\] 
If $x_{i, k} = 0$ for some $i$ and $k$, we add a small constant to it to evaluate the log ratio. For the initial value $\bmu^{(0)}$, we have a closed form minimizer of $\bmu$ for (\ref{objective.function}) given the values of $(\bz_1,\ldots,\bz_n)$, which is $\bmu = \bar{\bz} = \frac1n \sum_{i=1}^n \bz_i$. Therefore, we set the initial value as $\bmu^{(0)} = \frac1n \sum_{i=1}^n \bz_i^{(0)}$. Finally, for the initial value $\bOmega^{(0)}$, we use the estimate of the graphical lasso algorithm taking the sample covariance matrix computed from $\bz_1^{(0)},\ldots,\bz_n^{(0)}$ as input.

In step 1, given $\bmu^{(t)}$ and $\bOmega^{(t)}$, minimizing the objective function (\ref{objective.function}) with respect to $(\bz_1,\ldots,\bz_n)$ is equivalent to minimizing the following objective function with respect to each $\bz_i$ separately, for $i = 1,\ldots,n$:
\begin{equation}
\ell_i^{(t)}(\bz_i) = \frac12 (\bz_i - \bmu^{(t)})' \bOmega^{(t)} (\bz_i - \bmu^{(t)}) - \left[\sum_{k=1}^{K} x_{i,k} z_{i,k} - M_i \log(\sum_{k=1}^K e^{z_{i,k}} + 1)\right]. \label{objective.function.zi}
\end{equation}
The above objective function is a smooth and convex function in $\bz_i$ with the following positive definite Hessian matrix
\begin{equation*}
\bOmega^{(t)} + \frac{M_i}{\left(\sum_{k=1}^K e^{z_{i,k}} + 1\right)^2} \left\{\left(\sum_{k=1}^K e^{z_{i,k}} + 1\right) \text{diag} (e^{\bz_i}) - (e^{\bz_i})(e^{\bz_i})'\right\},
\end{equation*}
where $e^{\bz_i} = (e^{z_{i,1}}, \ldots, e^{z_{i,K}})'$ and $\text{diag}(e^{\bz_i})$ is the diagonal matrix with the diagonal elements $e^{z_{i,1}}, \ldots, e^{z_{i,K}}$. Therefore, we apply the Newton-Raphson algorithm to find the minimizer numerically. In addition, we implement a line search procedure in each Newton-Raphson iteration following the Armijo rule \citep{armijo1966minimization}. This procedure ensures sufficient decrease in the objective function at each iteration to prevent possible divergence of the algorithm.

Step 2 is similar to the initialization step, in which $\bmu$ has a closed-form solution and is updated as $\bar{\bz}^{(t+1)} = \frac1n \sum_{i=1}^n \bz_i^{(t+1)}$ from the current numerical values of $(\bz_1^{(t+1)},\ldots,\bz_n^{(t+1)})$ that are computed from the Newton-Raphson algorithm in step 1.

In step 3, given $(\bz_1^{(t+1)},\ldots,\bz_n^{(t+1)})$ and $\bmu^{(t+1)} = \bar{\bz}^{(t+1)}$, the objective function for $\bOmega$ can be simplified as
\begin{align}
\ell^{(t)}(\bOmega) &= -\frac12 \log[\det(\bOmega)] + \frac{1}{2n} \sum_{i=1}^n (\bz_i^{(t+1)} - \bmu^{(t+1)})' \bOmega (\bz_i^{(t+1)} - \bmu^{(t+1)}) + \lambda \|\bOmega\|_1, \notag \\
&= -\frac12 \log[\det(\bOmega)] + \frac{1}{2} \mathrm{tr} (\bS^{(t+1)} \bOmega) + \lambda \|\bOmega\|_1, \label{objective.function.omega} 
\end{align}
where $\bS^{(t+1)} = \frac{1}{n} \sum_{i=1}^n (\bz_i^{(t+1)} - \bar{\bz}^{(t+1)}) (\bz_i^{(t+1)} - \bar{\bz}^{(t+1)})'$. It is obvious that minimizing the objective function (\ref{objective.function.omega}) becomes a graphical lasso problem \citep{yuan2007model, banerjee2008model, friedman2008sparse}. It is well known that the graphical lasso objective function is a convex function in $\bOmega$ \citep{banerjee2008model} and efficient algorithms have been developed for its optimization \citep{friedman2008sparse}. In this paper, we implement this step using the graphical lasso algorithm included in the \texttt{huge} \citep{zhao2012huge} package in R.

The above block coordinate descent algorithm iterates between Newton-Raphson and graphical lasso and is designed specifically to optimize the objective function (\ref{objective.function}) for compositional count data. Therefore, we name this algorithm the compositional graphical lasso algorithm, and the entire approach the compositional graphical lasso method including both the model and the algorithm for the analysis of microbiome abundance data. 

\subsection{Theoretical Convergence}
\label{Sec-2.4}

Unfortunately, the objective function (\ref{objective.function}) is not necessarily a convex function jointly in $(\bz_1,\ldots,\bz_n)$, $\bmu$, and $\bOmega$. However, 
we have shown that it is convex in each of the three subsets of its parameters. The convergence property of such an optimization problem has been studied in the literature. For example, \cite{tseng2001convergence} studied the convergence property of a block coordinate descent method applied to minimize a nonconvex function with certain separability and regularity properties. We will establish the convergence property of the compositional graphical lasso algorithm following \cite{tseng2001convergence}.

Recall that our algorithm treats the three sets of parameters $(\bz_1,\ldots,\bz_n)$, $\bmu$, and $\bOmega$ as three blocks and optimizes for each block iteratively. In addition, as in \cite{tseng2001convergence}, the objective function (\ref{objective.function}) can be regarded as the sum of two parts, the first of which is an inseparable but differentiable function given by
\begin{equation}
\ell_0(\bz_1,\ldots,\bz_n,\bmu,\bOmega) = \frac{1}{2n} \sum_{i=1}^n (\bz_i - \bmu)' \bOmega (\bz_i - \bmu), \label{objective.function.inseparable}
\end{equation}
and the second of which is a sum of separable and differentiable functions given by $\ell_1(\bz_1,\ldots,\bz_n) + \ell_2(\bOmega)$, where
\begin{align}
\ell_1(\bz_1,\ldots,\bz_n) &= -\frac1n\sum_{i=1}^n \left[\sum_{k=1}^{K} x_{i,k} z_{i,k} - M_i \log(\sum_{k=1}^K e^{z_{i,k}} + 1)\right], \label{objective.function.separable.1}\\
\ell_2(\bOmega) &= -\frac12 \log[\det(\bOmega)] + \lambda \|\bOmega\|_1. \label{objective.function.separable.2}
\end{align} 
\cite{tseng2001convergence} established the convergence property of a block coordinate descent algorithm under regularity conditions on $\ell_0$, $\ell_1$, and $\ell_2$.

To present the main convergence property of the compositional graphical lasso algorithm, let's review the definition of a cluster point in real analysis. A cluster point of a set $\mathcal{A} \subset \mathbb{R}^n$ is a real vector $\mathbf{a} \in \mathbb{R}^n$ such that for every $\delta > 0$, there exists a point $\mathbf{x}$ in $\mathcal{A}\setminus \{\mathbf{a}\}$ satisfying that $\|\mathbf{x} - \mathbf{a}\|_2 < \delta$. Obviously, any limit point of the set $\mathcal{A}$ is a cluster point. Furthermore, define a cluster point of the compositional graphical lasso algorithm to be a cluster point of the set $\{(\bz_1^{(t)},\ldots,\bz_n^{(t)},\bmu^{(t)},\bOmega^{(t)}): t = 0,1,2,\ldots\}$, which are minimizers found at each iteration $t$. Then, the following theorem presents a theoretical property for every cluster point of our algorithm as follows.

\begin{thm} \label{Thm-1}
Any cluster point of the compositional graphical lasso algorithm is a stationary point of the objective function (\ref{objective.function}).
\end{thm}

The proof of Theorem \ref{Thm-1} can be found in the supplementary materials. This theorem guarantees that a cluster point, usually a limit point, of the compositional graphical lasso algorithm, is at least a stationary point. It is noteworthy that there exists a global minimizer for the objective function in (\ref{objective.function}) because we have proved the coerciveness of the objective function in the proof \citep[Lemma 8.4]{calafiore2014optimization}. Therefore, to achieve global optimization in practice, one can run the algorithm multiple times starting with different initial values and choose the one solution that yields the smallest objective function among the multiple ones.

In addition, the values of the objective function at each iteration, i.e., \\
$\{\ell(\bz_1^{(t)},\ldots,\bz_n^{(t)},\bmu^{(t)},\bOmega^{(t)}): t = 0,1,2,\ldots\}$, will always converge. This is because that the objective function is bounded below (as $\ell_0 + \ell_2$ is bounded below as shown in the proof and $\ell_1 > 0$ by definition)
and our algorithm results in non-increasing objective function values between two iterations. Therefore, the values of objective function will always converge to a limit point. In practice, we have always observed the numerical convergence of both the minimizers and the values of the objective function after a certain number of iterations.



\subsection{Tuning Parameter Selection}
\label{Sec-2.5}

There is a large body of literature on the selection of a tuning parameter in the variable selection framework. Common approaches can be broadly categorized into three types: criteirion-based methods, prediction-based methods, and stability-based methods. Criterion-based methods such as Akaike information criterion (AIC) \citep{akaike1974new} and Bayesian information criteria (BIC) \citep{schwarz1978estimating} balance the model complexity and the goodness of fit, prediction-based methods such as cross validation \citep{stone1974cross, geisser1975predictive} and generalized cross validation \citep{golub1979generalized} aim to minimize the expected prediction error of the selected model on independent datasets. Stability-based methods such as stability selection \citep{meinshausen2010stability} and the Stability Approach to Regularization Selection (StARS) \citep{liu2010stability} select a model with high stability under subsampling or bootstrapping the original data.

In this work, we apply StARS to select the tuning parameter $\lambda$ in our objective function (\ref{objective.function}). In StARS, we draw $N$ subsamples without replacement from the original dataset with $n$ observations, each subsample of size $b$. For each value of the tuning parameter $\lambda$, we obtain an estimate of $\bOmega$, i.e., a network for each subsample. Then, we measure the total instability of these resultant networks across the $N$ subsamples. The total instability of these networks is defined by averaging the instabilities of each edge across the $N$ subsamples over all possible edges, where the instability of each edge is estimated as the twice the sample variance of the Bernoulli indicator of whether this edge is selected or not in each of the $N$ subsamples.

Starting from a large penalty which corresponds to the empty network, the instability of networks increases as $\lambda$ decreases. StARS stops and selects the tuning parameter to be the minimum value of $\lambda$'s with which the instability of the resultant networks is less than a threshold $\beta > 0$. In principle, StARS selects a tuning parameter so that the resultant network is the densest among networks with a total instability less than a threshold $\beta$ without violating some sparsity assumption. The selected network is the ``densest on the sparse side," as it starts with the empty network and stops when the instability first across the threshold. 

\section{Simulation Studies}
\label{Sec-3}

\subsection{Simulation Settings}
\label{Sec-3.1}

To evaluate the performance of compositional graphical lasso, we conduct \textcolor{blue}{simulation studies} and compare it with other network estimation methods. 

Given that our goal is to estimate the true network, i.e., $\bOmega$ in (\ref{logistic.normal}), we consider the following three types of true precision matrices $\bOmega = (\omega_{kl})_{1\le k,l \le K}$, which are different in the pattern of edge distributions as well as the degree of connectedness.

\begin{enumerate}[nosep]
\item Chain network: $\omega_{kk} = 1.5$, $\omega_{kl} = 0.5$ if $|k - l| = 1$, and $\omega_{kl} = 0$ if $|k - l| > 1$. A node is designed to be connected to its adjacent nodes, and the connectedness of nodes is balanced.
\item Random network: $\omega_{kl} = 1$ with probability $3/K$ for $k \ne l$. A node is connected to all other nodes randomly with a fixed probability, set to be $3/K$. Similar to the chain structure, the connectedness of nodes is balanced.
\item Hub network: All nodes are randomly split into $\lceil K / 20 \rceil$ disjoint groups, and a hub node $k$ is selected from each group. For any other node $l$ in the same group, $\omega_{kl} = 1$. All the remaining entries of $\bOmega$ are zero. Here, nodes are partitioned into the same group at random, but is then designated to be connected to the hub node at certain. The degree of connectedness among nodes is extremely unbalanced in this case: the hub nodes are connected to all the other nodes in its group (around 20 nodes) and all the other nodes are only connected to the hub node in its group, i.e., just one node.
\end{enumerate}

In addition to the true network, we also consider two other factors that are expected to influence the result. The first factor is the sequencing depth, $M_i$, in the multinomial distribution (\ref{multinomial}). \textcolor{blue}{We simulate $M_i$ from uniform distributions, the detail of which will be discussed in the following subsections.} The second factor is the degree of variation in the logistic normal distribution (\ref{logistic.normal}). For each of the three types of precision matrices, we consider an additional factor by multiplying a positive constant $c$ to $\bOmega$ so that the true precision matrix is $c\bOmega$. We choose $c = 1$ and $c = 1/5$ separately and call the two settings low and high compositional variation, respectively.

The data are simulated following the logistic normal multinomial model in (\ref{multinomial})--(\ref{logistic.normal}). We first simulate $\bz_i \sim N (\bmu, \bSigma)$ independently for $i = 1,\ldots,n$; then, we perform the inverse log-ratio transformation (also know as the softmax transformation, the inverse transformation of (\ref{log.ratio.transformation})) to obtain the multinomial probabilities $\bp_i$ for $i = 1,\ldots,n$; last, we simulate multinomial counts $\bx_i$ from a multinomial distribution with sequencing depth $M_i$ and probabilities $\bp_i$. Throughout this simulation study, we fix $n = 100$ and $K = 200$.

The simulation results are based on 100 replicates. For each replicate, we apply compositional graphical lasso, neighborhood selection, and graphical lasso separately to obtain a sparse estimator of $\bOmega$. For neighborhood selection and graphical lasso, we first obtain an estimate of $\bz_1,\ldots,\bz_n$ from the multinomial distribution via the additive log-ratio transformation
\begin{equation}
\tilde{z}_{i,k}= \log (\frac{x_{i, k}}{x_{i, K + 1}}),\ i = 1,\ldots,n, \ k = 1,\ldots,K, \label{z.surrogates}
\end{equation}
and then apply neighborhood selection and graphical lasso directly on the estimates $\tilde\bz_1,\ldots,\tilde\bz_n$ by treating them as surrogates for their true counterparts, i.e., $\bz_1,\ldots,\bz_n$. These methods are almost identical to SPIEC-EASI although we replaced the centered log-ratio transformation in SPIEC-EASI with additive log-ratio transformation for a fair comparison.

To compare the performance of the three methods in terms of network recovery, all three methods are applied with a sequence of tuning parameter values, and their true positive rates (TPR) and false positive rates (FPR) in terms of edge selection are recorded for each value of $\lambda$. An ROC curve is plotted from the average TPR and the average FPR over the 100 replicates for each of the tuning parameters.

In addition, we apply StARS to select an optimal tuning parameter $\lambda$. Following the recommendation in \citet{liu2010stability}, we set the threshold for the total instability to be $\beta = 0.05$, the size of each subsample $b = 7 \sqrt{n}$, and the number of subsamples $N = 50$. Once the optimal tuning parameter is determined by StARS, we fit the whole dataset with the selected tuning parameter and evaluate the resultant network using three criteria: precision, recall, and F1 score, which are defined as
\begin{equation*}
\text{Precision} = \frac{\text{TP}}{\text{TP} + \text{FP}},\quad \text{Recall} = \frac{\text{TP}}{\text{TP} + \text{FN}}, \quad \text{F1} = \frac{2 \times \text{Precision} \times \text{Recall}}{\text{Precision} + \text{Recall}},
\end{equation*}
where TP, FP, and FN are numbers of true positives, false positives, and false negatives, respectively. 

\subsection{\textcolor{blue}{Simulation Results for Dense Data}}
\label{Sec-3.2}

\textcolor{blue}{In this subsection, we evaluate the performance of compositional graphical lasso on dense data, in which most of the simulated counts are nonzero. This corresponds to a simulation setting where the sequencing depths $M_i$'s are relatively larger than the number of taxa $K + 1$. Still, to evaluate the effect of the sequencing depth on the performance of network estimation methods, we simulate $M_i$ from two uniform distributions, Uniform$(20K, 40K)$ and Uniform$(100K, 200K)$, and call the two settings low and high sequencing depth in this subsection, respectively.}

\begin{figure}[h]
\centering
\includegraphics[width=0.8\textwidth]{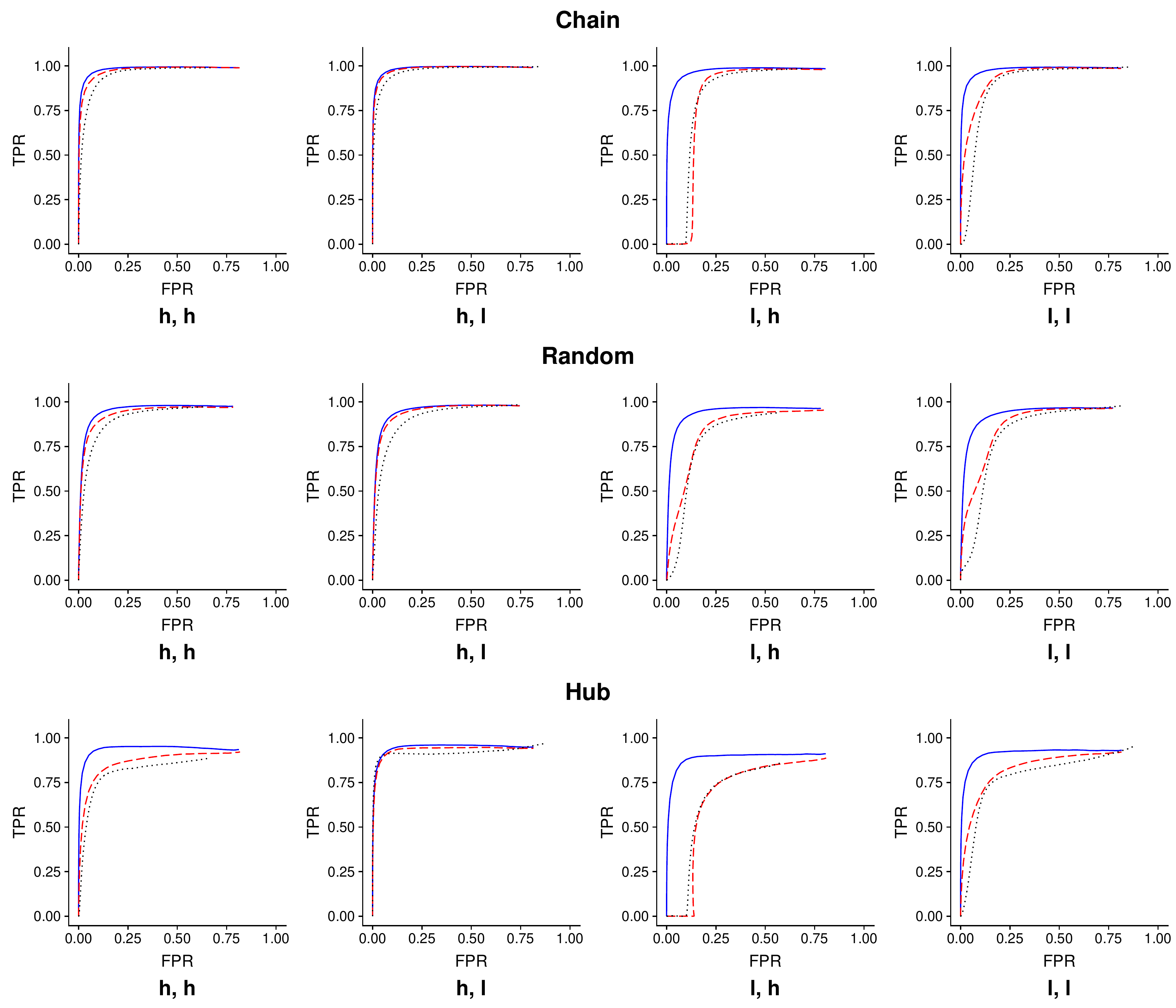}
\caption{ROC curves for compositional graphical lasso (Comp-gLASSO), graphical lasso (gLASSO) and neighborhood selection (MB). Solid blue: Comp-gLASSO; dashed red: gLASSO; dotted black: MB. $\mathbf{h/l, h/l}$: high/low sequencing depth, high/low compositional variation.}
\label{ROC}
\end{figure}

Figure \ref{ROC} presents the ROC curves for compositional graphical lasso (Comp-gLASSO), neighborhood selection (MB), and graphical lasso (gLASSO), from which we can see that compositional graphical lasso dominates its competitors in terms of edge selection in all settings. In particular, the advantage of the compositional graphical lasso over neighborhood selection and graphical lasso is the most obvious when the compositional variation is high and the sequencing depth is low, no matter which type of network structure is considered. On the contrary, the three methods perform very similarly for all types of network structures when the compositional variation is low and the sequencing depth is high. The difference between compositional graphical lasso and the rest is intermediate for the other two settings when both compositional variation and sequencing depth are high or when both are low. Comparing graphical lasso and neighborhood selection, they tend to perform more similarly although graphical lasso seems to outperform neighborhood selection in some settings with a small margin.

\begin{figure}[h]
\centering
\includegraphics[width=0.8\textwidth]{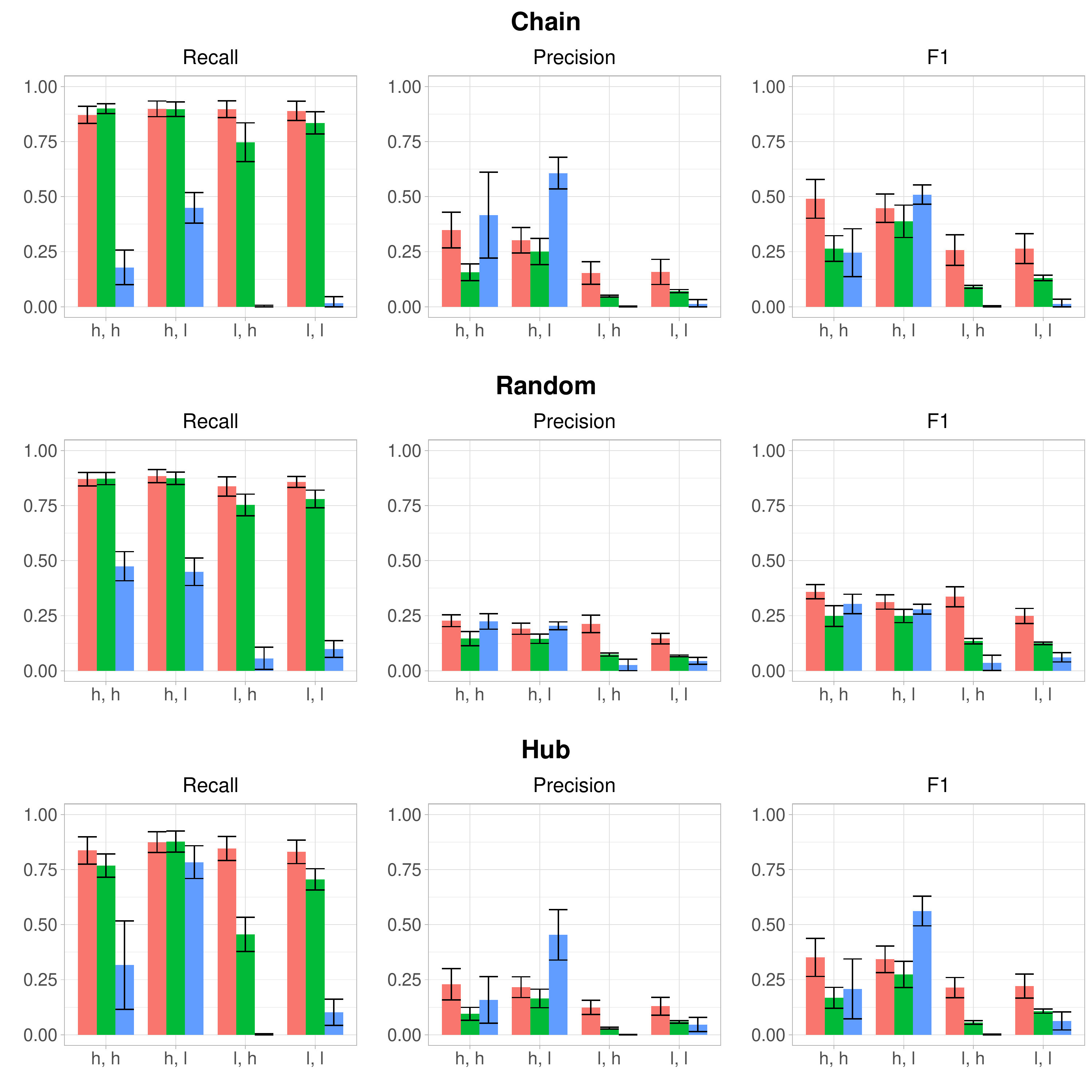}
\caption{Recall, precision and F1 score for the network selected by StARS for compositional graphical lasso (Comp-gLASSO), graphical lasso (gLASSO) and neighborhood selection (MB). Red (left): Comp-gLASSO; green (middle): gLASSO; blue (right): MB. $\mathbf{h/l, h/l}$: high/low sequencing depth, high/low compositional variation.}
\label{Recall.Precision.F1}
\end{figure}

The above observations agree with our expectation about how the two factors, compositional variation and sequencing depth, affect the comparison between the methods. Recall that neighborhood selection and graphical lasso replace the true values of $\bz_1,\ldots,\bz_n$ by their estimates/surrogates $\tilde\bz_1,\ldots,\tilde\bz_n$ as in (\ref{z.surrogates}) without taking into account the estimation accuracy or uncertainty of these surrogates. First, a higher sequencing depth leads to more accurate surrogates $\tilde\bz_1,\ldots,\tilde\bz_n$; therefore, it is not surprising to see that the three methods perform more similarly when the sequencing depth is high. Second, a higher compositional variation results in a higher variation in $\bz_i$'s and further in $\bp_i$'s that are multinomial probabilities. Since neighborhood selection and graphical lasso ignore the multinomial component in the model, it is also not surprising to see that their performance is deteriorated by a high compositional variation.

Figure \ref{Recall.Precision.F1} presents the recall, precision, and F1 score from 50 replicates of the estimated network resulted from the tuning parameter selected by StARS. The first observation would be that the precisions of both compositional graphical lasso and graphical lasso are much worse than their recalls, whereas the precisions and recalls are more comparable for neighborhood selection. Interestingly, StARS results in a much more sparse network for neighborhood selection than the other methods under the same stability threshold, suggesting that fewer edges selected by neighborhood selection are stable enough (within the instability threshold). When it comes to method comparison, compositional graphical lasso has much higher recall than neighborhood selection in most settings, but have comparable or lower precision in most of the settings with high sequencing depth. The network from compositional graphical lasso has higher F1 score than the ones from neighborhood selection in most settings, except when sequencing depth is high and compositional variation is low for chain and hub networks. In addition, the network from compositional graphical lasso has higher or comparable precision, recall, and F1 score than the ones from graphical lasso in all settings. Similar to the observations from the ROC curves, the advantage of compositional graphical lasso is more obvious with a low sequencing depth or a high compositional variation.

\subsection{\textcolor{blue}{Simulation Results for Sparse Data}}
\label{Sec-3.3}

\textcolor{blue}{In this subsection, we evaluate the performance of compositional graphical lasso on sparse data, in which a range of proportions of the compositional counts are simulated as zero. We only present the simulation results for the chain network, given that the simulation results for the other network types are similar. To examine how our method performs for different sparsity levels, we simulate $M_i$ from four uniform distributions, Uniform$(8K, 16K)$, Uniform$(4K, 8K)$, Uniform$(2K, 4K)$, and Uniform$(K, 2K)$, respectively. Note that the sparsity level of the simulated data also depends on the compositional variation (see Section \ref{Sec-3.1}). In a typically simulated dataset with high compositional variation, the sparsity level is around 40\%, 50\%, 60\%, and 70\% with the four uniform distributions for $M_i$; in a typically simulated dataset with low compositional variation, the sparsity level is around 5\%, 10\%, 25\%, and 40\% with the four uniform distributions for $M_i$. In both cases, we refer to these four sparsity levels as 1, 2, 3, and 4, with 1 the least sparse and 4 the most sparse. Due to the limited space, we place the figures resulted from this simulation in the supplementary materials.}

\textcolor{blue}{Figure \ref{ROC_MB_Sparse} presents the ROC curves for compositional graphical lasso (Comp-gLASSO), neighborhood selection (MB), and graphical lasso (gLASSO). Similar to Figure \ref{ROC}, we can see that compositional graphical lasso dominates its competitors in terms of edge selection across all sparsity levels. As the sequencing depths for sparse data are relatively small compared to the ones used for dense data, the advantage of the compositional graphical lasso over neighborhood selection and graphical lasso is obviously and consistently observed in all simulation settings. In addition, graphical lasso and neighborhood selection tend to perform more similarly although graphical lasso seems to outperform neighborhood selection with a small margin. This result is consistent with that observed for dense data in Section \ref{Sec-3.2}.}

\textcolor{blue}{Figure \ref{Recall.Precision.F1.Sparse} presents the recall, precision, and F1 score from 50 replicates of the estimated network resulted from the tuning parameter selected by StARS. The first observation is that all methods perform worse for sparse data than for dense data from the comparison between Figure \ref{Recall.Precision.F1.Sparse} and Figure \ref{Recall.Precision.F1}; all methods perform worse as the sparsity level increases. Similar to Figure \ref{Recall.Precision.F1}, Figure \ref{Recall.Precision.F1.Sparse} implies that compositional graphical lasso outperforms graphical lasso and neighborhood selection, with a higher recall, precision, and F1 score consistently observed in most settings. The only exception is that, when data are extremely sparse with a low compositional variation, graphical lasso seems to outperform compositional graphical lasso by a very small margin as both methods are not working well. For sparse data, StARS results in an almost empty network for neighborhood selection, resulting in zero recall, precision, and F1 score for most settings. This agrees with our observation from the simulation results for dense data that neighborhood selection tends to produce a much sparser network than compositional graphical lasso and graphical lasso when its tuning parameter is selected by StARS.}

\section{Real Data}
\label{Sec-4}

\subsection{Benchmark Study: \textit{Tara} Oceans Project}
\label{Sec-4.1}
To better understand the ocean, the largest ecosystem on the earth, the \textit{Tara} Oceans Project aims to build the global ocean interactome that can be used to predict the dynamics and structure of ocean ecosystems. To achieve this, the \textit{Tara} Oceans Consortium sampled both plankton and environmental data at 210 sites from the world's oceans using the 110-foot research schooner \textit{Tara} during the \textit{Tara} Oceans Expedition (2009-2013). The data collected was later processed using sequencing and imaging techniques. One unique advantage of the \textit{Tara} Oceans Project is that it has generated a list of 91 genus-level marine planktonic interactions that have been validated in the literature \citep{lima2015determinants}. Though this list only comprises interactions between microbes that represent a small fraction of the total marine eukaryotic diversity and is therefore far from complete, it could serve as partial ground truth for us to evaluate the interactions identified by different methods. Thus, our major goal is to use the \textit{Tara} Oceans Project as a benchmark study in order to compare the performance of different methods in constructing the planktonic interactions.

As the partial ground truth is a list of genus-level interactions, we choose to analyze the genus-level abundance data, which are aggregated from the original OTU abundance data downloadable from the \textit{Tara} Oceans Project data repository (\url{http://taraoceans.sb-roscoff.fr/EukDiv/}). As a benchmark study, we only include the 81 genera that are involved in the list of gold-standard interactions in our analysis. In addition, we discard the samples with too few sequence reads (less than 100), resulting in 324 samples left in our analysis. 

Similar to the simulation study, we apply compositional graphical lasso, graphical lasso, and neighborhood selection to estimate the interaction network among the 81 genera. We first pick the genus \textit{Acrosphaera}, which has the largest average relative abundance among those genera not involved in the gold-standard list, and use this genus as the reference taxon for all three methods. Then, we apply each method with a sequence of 70 decreasing tuning parameter values, resulting in a sequence of interaction networks starting from an empty network. Finally, we apply StARS to find the optimal tuning parameter, in which the parameters $\beta$, $b$, and $N$ are set the same as in the simulation study.

\begin{figure}[htbp]
\centering
\begin{subfigure}{.4\textwidth}
  \centering
  \includegraphics[width=1\linewidth]{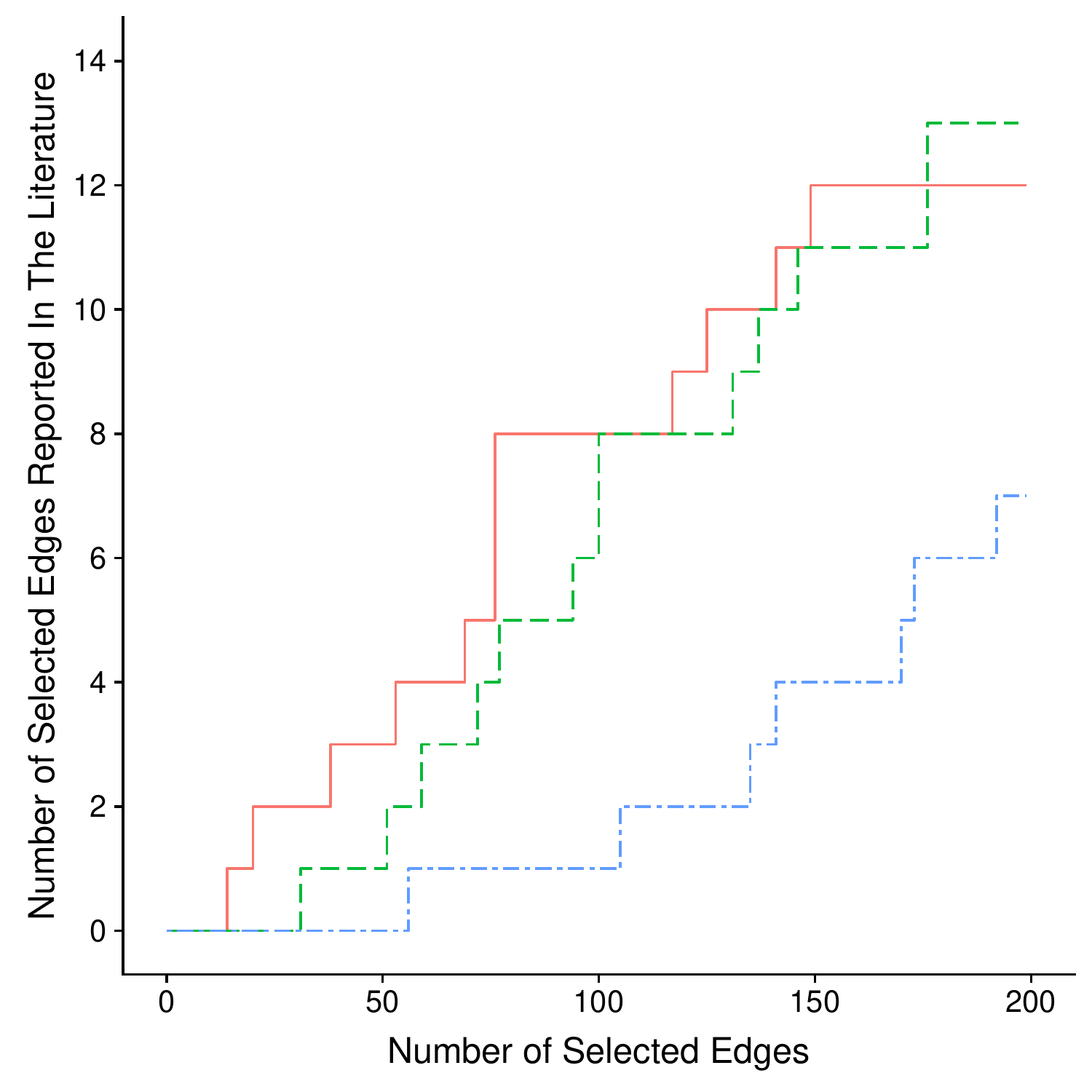}
  \caption{}
  \label{tara_path}
\end{subfigure}%
\begin{subfigure}{.4\textwidth}
  \centering
  \includegraphics[width=1\linewidth]{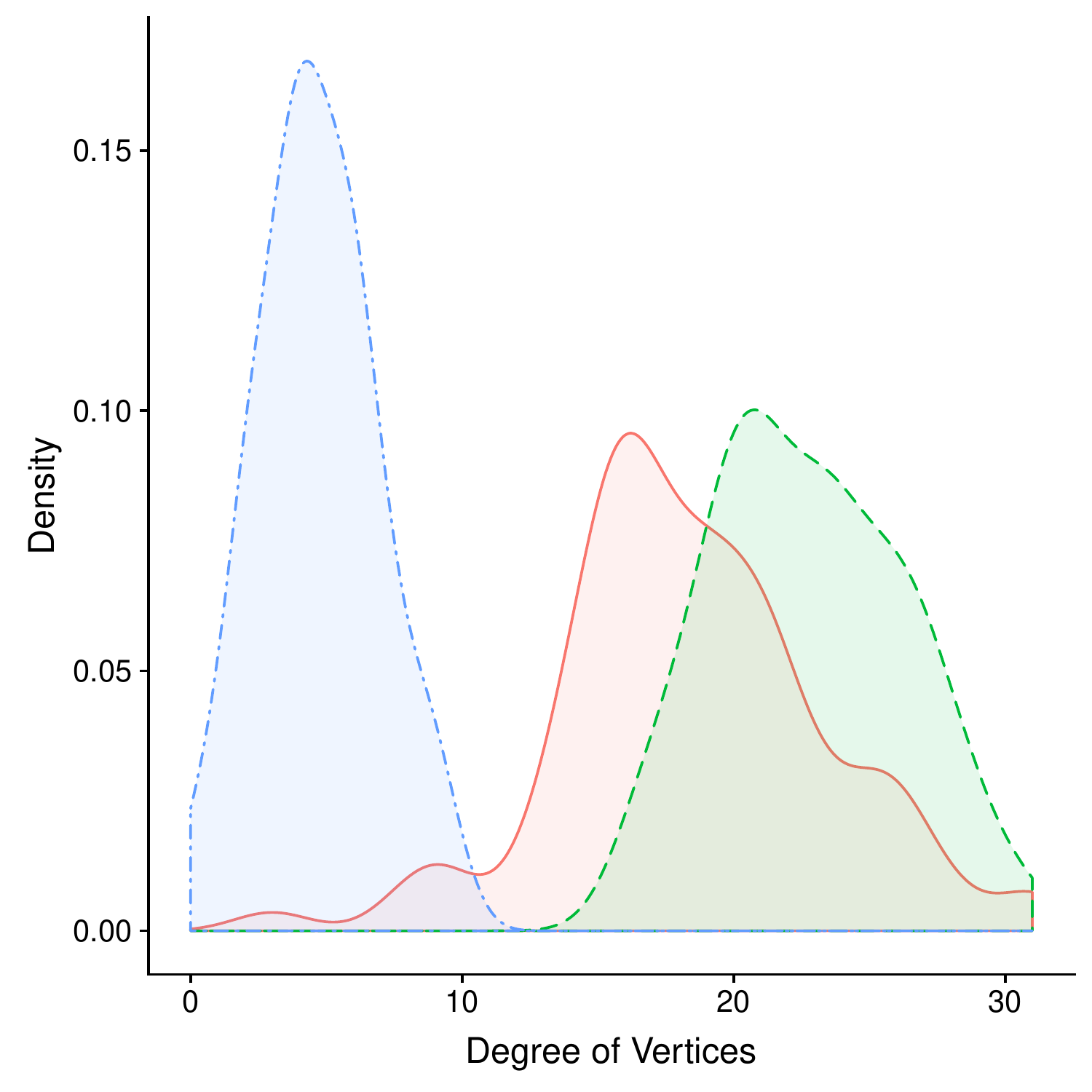}
  \caption{}
  \label{tara_degree}
\end{subfigure}
\caption{(a): Number of identified literature interactions versus number of edges of the estimated network from the \textit{Tara} dataset. (b): The degree distribution of vertices from the networks selected by StARS. Solid red: compositional graphical lasso; dashed green: graphical lasso; dashed dotted blue: neighborhood selection.}
\label{network.P}
\end{figure}

First, we compare the three methods in terms of how highly they rank the literature validated interactions amongst their top reported edges. Specifically, we start with a large value of the tuning parameter that results in an empty network, then decrease the tuning parameter so that the network becomes denser, and stop until the network has about 200 edges (out of a total of 3240 possible edges). At each value of the tuning parameter, we plot the number of literature validated interactions included in the network versus the total number of edges of the network, resulting in a step function for each method (Figure \ref{tara_path}). From Figure \ref{tara_path}, we observe that compositional graphical lasso identifies slightly more literature validated interactions than graphical lasso until the total number of edges arrives 175 and graphical lasso identifies one more literature validated interaction afterwards. Neighborhood selection selects much fewer literature validated interactions than either compositional graphical lasso or graphical lasso. These observations imply that compositional graphical lasso slightly outperforms graphical lasso in reconstructing the literature validated interactions, while its advantage over neighborhood selection is much more obvious.

Second, we compare the overall topologies of the three interaction networks with tuning parameters selected by StARS. We find that compositional graphical lasso, graphical lasso, and neighborhood selection identify 749, 921, and 190 edges, respectively, with the same instability threshold used in StARS. This agrees with our observation in the simulation study that the network from neighborhood selection is much sparser than those from compositional graphical lasso and graphical lasso. The degree distributions from the networks estimated by the three methods are shown in Figure \ref{tara_degree}. 
The center of the three degree distributions are ranked as neighborhood selection, compositional graphical lasso, and graphical lasso in the ascending order, which is also reflected in the densities of the three interaction networks.

Third, we investigate the high-degree nodes, i.e., hub genera, in the interaction networks. High-degree nodes are often thought to represent taxa that elicit an effect on a large number of other members of their community, such as keystone taxa and generalistic parasites. It is observed that there are a few hub genera that have an excessive number of interactions with other genera in the literature, such as \textit{Amoebophrya}, \textit{Blastodinium}, and \textit{Parvilucifera}, which are referred to as benchmark hub genera. Although the literature validated interactions are rather incomplete, it is still of interest to evaluate how well the three methods pick up those benchmark hub genera. Since the density of networks from the three methods are rather different, it is hard to compare the degrees of the hub genera from the three networks directly, but it is reasonable to compare the ranks of those degrees within each degree distribution. The method that generates lower ranks (degree of genera ranked in descending order) for those hubs in their degree distribution are believed to pick up the benchmark hub genera better. 

A list of 7 benchmark hubs (which has degree $\geq 5$) along with their degrees from the incomplete network constructed from the literature is shown in Table \ref{tab:degree rank of lit hubs}, followed by the corresponding ranks of those genera in the degree distributions from each of the three methods and their corresponding degrees in the parentheses. We can see that compositional graphical lasso generates lower ranks than graphical lasso for all 7 genera, while neighborhood selection generates lower ranks than compositional graphical lasso for 3 genera, and the opposite for the other 4 genera. Overall, compositional graphical lasso performs the best in picking up the benchmark hub genera among the three methods.

\begin{table}[htbp]
\centering
\begin{tabular}{lllll}
  \hline
 & Literature & Comp-gLASSO & gLASSO & MB \\ 
  \hline
  \textit{Amoebophrya} & 1 (21) & 31 (19) & 47 (21) & 2 (9) \\ 
  \textit{Blastodinium} & 2 (12) & 13 (23) & 26 (24) & 30 (5) \\ 
  \textit{Parvilucifera} & 2 (12) & 46 (17) & 67 (19) & 58 (3) \\ 
  \textit{Syndinium} & 4 (7) & 14 (23) & 27 (24) & 19 (6) \\ 
  \textit{Vampyrophrya} & 4 (7) & 34 (19) & 60 (20) & 6 (8) \\ 
  \textit{Phaeocystis} & 6 (6) & 1 (31) & 3 (29) & 17 (6) \\ 
  \textit{Pirsonia} & 7 (5) & 64 (15) & 68 (19) & 33 (5) \\
   \hline
\end{tabular}
\caption{\label{tab:degree rank of lit hubs} For the hub genera, their ranks in each degree distribution (in descending order) from the literature, compositional graphical lasso (Comp-gLASSO), graphical lasso (gLASSO) and neighborhood selection (MB). The numbers in the parentheses are the corresponding degrees of the genera.}
\end{table}

We find that compositional graphical lasso predicts a high degree for genera that are known to act as keystone species or parasitize other taxa, many of which are also high-degree in the literature validated network. For example, \textit{Phaeocystis} elicits a high degree in both literature validated and compositional graphical lasso networks. This genus is a well-described keystone organism in some marine ecosystems \citep{verity2007current}, where it causes large phytoplankton blooms and plays an important role in the global cycling of carbon and sulfur \citep{verity1996organism}. Similarly, these two networks predict a high degree for taxa that are known parasites, such as \textit{Blastodinium} \citep{skovgaard2012parasitic}, \textit{Amoebophyra} \citep{chambouvet2008control}, and \textit{Syndinium} \citep{skovgaard2005phylogenetic}. In some instances, compositional graphical lasso uniquely reveals high-degree genera even when compared to the literature validated network, such as in the case of the parasite \textit{Euduboscquella} \citep{bachvaroff2012molecular}, which is ranked the 5th by compositional graphical lasso but only has 1 interaction in the literature. Given the fact that the literature validated network only captures a subset of interactions that exist in nature (i.e., those interactions that have been explicitly tested), we posit that compositional graphical lasso affords an opportunity to resolve novel modulators of community composition, such as keystone taxa and generalistic parasites, that follow-up experiments can validate.

\subsection{Zebrafish Parasite Infection Study}
\label{Sec-4.2}
The Zebrafish Parasite Infection Study, recently conducted at Oregon State University, used a zebrafish helminth intestinal infection model to resolve how the gut microbiome and parasite burden covary during infection \citep{gaulke2019longitudinal}. This study quantified the infection burden of an intestinal helminth of zebrafish, \textit{Pseudocapillaria tomentosa}, and the gut microbiome in 210 4-month-old zebrafish. Half of these fish were exposed to the parasite, \textit{P.\ tomentosa}, and the other half were unexposed. Given that not all exposed fish would be ultimately infected, parasite burden was measured more accurately as the total number of worms in their fecal samples. In addition, the same fecal samples were used to measure the abundance of the gut microbiome species via 16S rRNA sequencing, which resulted in gut microbiome data for 207 fish. Among these fish, 81 were infected after being exposed to \textit{P.\ tomentosa} and the other 126 fish were not infected, as indicated by the total number of worms. Our major goal is to evaluate how the microbial interaction network associates with successful parasite infection in the gut.

Similar to the \textit{Tara} Oceans Project, we choose to analyze the genus-level abundance data. We discard those genera that have a nonzero abundance in less than 5\% of the samples, resulting in 42 genera left in our analysis. In addition, we follow the same strategy to define a reference genus as in \citet{jiang2020microbial}, i.e., combining those OTUs that do not have a genus-level taxonomic classification into a pseudo-genus and using it as the reference genus. The analysis is conducted in the same way as in the \textit{Tara} Oceans Project, except that the methods (compositional graphical lasso, graphical lasso, and neighborhood selection) are applied separately for the uninfected and the infected fish, resulting in three interaction networks for each group of fish. Analyzing the uninfected and infected fish separately allows us to compare the microbial interaction networks between the two groups.

\begin{figure}[htbp]
\centering
\begin{subfigure}{.4\textwidth}
  \centering
  \includegraphics[width=1\linewidth]{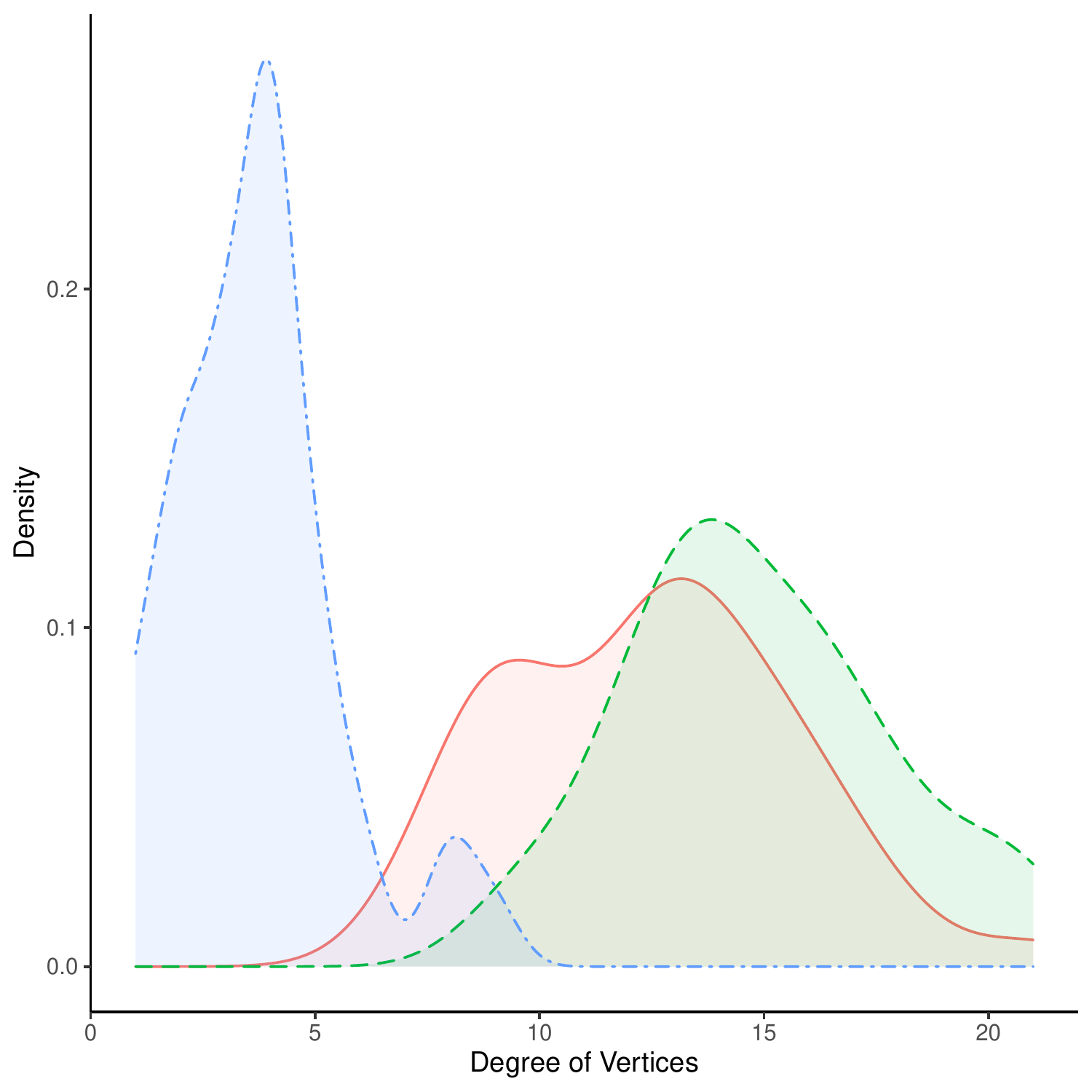}
  \caption{}
  \label{not_infected_degree}
\end{subfigure}%
\begin{subfigure}{.4\textwidth}
  \centering
  \includegraphics[width=1\linewidth]{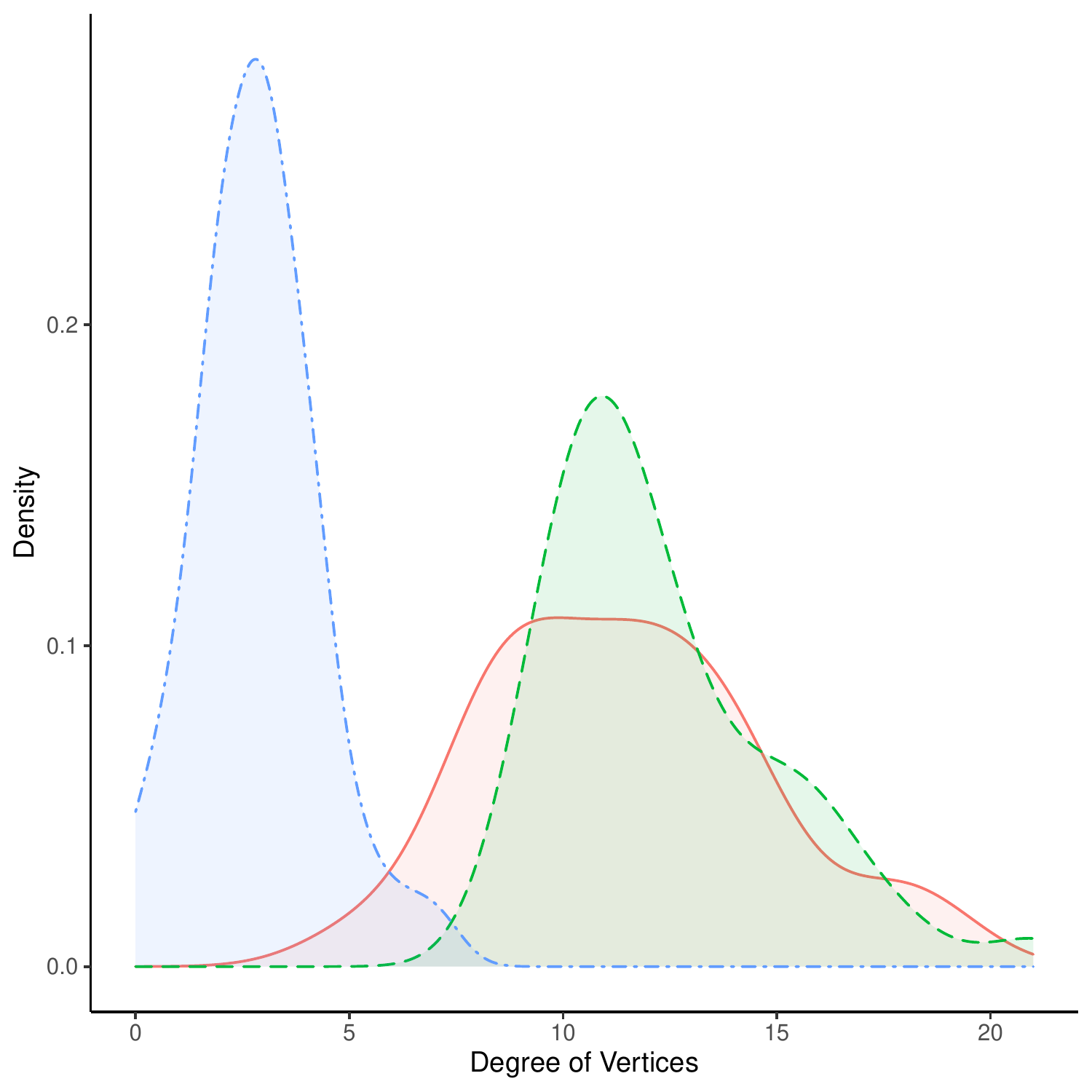}
  \caption{}
  \label{infected_degree}
\end{subfigure}
\caption{(a): The degree distribution of vertices from the networks selected by StARS for uninfected fish. (b): The degree distribution of vertices from the networks selected by StARS for infected fish. Solid red: compositional graphical lasso; dashed green: graphical lasso; dashed dotted blue: neighborhood selection.}
\label{fish_degree}
\end{figure}

First, we compare the overall topologies of the interaction networks with tuning parameters selected by StARS. We find that compositional graphical lasso, graphical lasso, and neighborhood selection identify 262, 312, and 78 edges, respectively, for uninfected fish, and 241, 259, and 60 edges, respectively, for infected fish. Comparing the two groups of fish, the interaction networks for the uninfected fish are slightly denser than the infected fish. The comparison among the three methods agrees with our observation in the simulation study and in the \textit{Tara} Oceans Project that the network from neighborhood selection tends to be much sparser than those from compositional graphical lasso and graphical lasso. The degree distributions based on the three methods are shown in Figure \ref{fish_degree}, which again suggests high similarity between the two groups of fish. Similar to the \textit{Tara} Oceans Project, neighborhood selection results in the lowest median degree, followed by composition graphical lasso and graphical lasso, regardless of whether the fish are infected or not.

Second, we further investigate the degree of each node in the interaction networks. This analysis helps identify those high-degree nodes, i.e., hub genera, which are indicative of keystone taxa in the microbial community. {\color{blue}Due to the limited space, we refer to Table \ref{fish_hubs} in the supplementary materials that presents all 42 genera with their degrees in each network in descending order.} Although the networks are similar in density between the two groups of fish, we have observed a substantial difference between the degrees of individual nodes. To see this we sort the nodes in each network based on their degrees and compare the top ten nodes between networks. Taking compositional graphical lasso as an example, only three out of the top ten nodes overlap between uninfected and infected fish: \textit{Paicibacter}, \textit{Photobacterium}, \textit{Fusobacterium}. This suggests that the network structures between the uninfected and infected fish are more distinctive than what their overall topologies might appear. The hub genera identified by different methods are quite different from each other too. For the group of uninfected fish, only three out of the top ten nodes overlap between the networks from the three methods: \textit{Aeromonas}, \textit{Photobacterium}, and \textit{Rheinheimera}; similarly, for the group of infected fish, only three out of the top ten nodes overlap between the networks from the three methods: \textit{Fusobacterium}, \textit{Paucibacter}, and \textit{Yersinia}.

{\color{blue}Third, we compare the networks generated from uninfected and infected fish to clarify the relationship between infection and the interactions that exist in the gut microbiome, motivated by the hypothesis that intestinal dysbiosis is defined not only by changes in community composition, but also by how members of the community ecologically interact. In order to understand this relationship, we identify microbes whose interaction set changes between infected and uninfected hosts then compare these results across methods. One of the bacterial taxa that compositional graphical lasso identifies as increasing in interaction degree among infected individuals, while other methods estimate a decrease in detected interaction degree, is the genus \textit{Gemmobacter}. Our earlier work \citep{gaulke2019longitudinal} showed that the abundance of this taxon positively links to parasite exposure, indicating that it is more ecologically successful in infected versus uninfected intestines. The analysis from compositional graphical lasso indicates that this increase in \textit{Gemmobacter} relative abundance in infected individuals is coincident with an increase in the ecological interactions between \textit{Gemmobacter} and other members of the gut microbiome, as it interacts with a greater number of other taxa in infected individuals. While \textit{Gemmobacter} is not a particularly well studied genus, prior work links it to dysbiotic diseases in a variety of hosts \citep{bates2022microbiome, ni2020gut}. Our observation of an increase in its interaction with other taxa in infected individuals strengthens the hypothesis that \textit{Gemmobacter} may act as an agent of dysbiosis, as its increase in its relative abundance not only links to a disease context, but its more integrated role in the microbial community (i.e., it potentially impacts a larger number of taxa in the gut).

The results from compositional graphical lasso also uniquely suggest that the \textit{Paucibacter} genus in uninfected individuals displays a higher number of identified interactions relative to infected individuals. The other methods we evaluated find that \textit{Paucibacter}'s interactions are relatively consistent between infected and uninfected individuals. The results from compositional graphical lasso are notable because the genus \textit{Paucibacter} includes clades that have been identified as being conserved in the healthy zebrafish gut microbiome \citep{sharpton2021phylogenetic}, presumably because these taxa are critical to the commensal microbial community. Based on our observations, namely that disruption to the broader microbiome by parasite infection appears to disrupt the set of interactions between \textit{Paucibacter} and other taxa in the community, we hypothesize that \textit{Paucibacter} is a keystone member of the healthy zebrafish gut microbiome and that parasite exposure induced dysbiosis occurs in part by disrupting \textit{Paucibacter} in the gut. Collectively, our observations reveal patterns of interaction between infected and uninfected individuals that clarify the potential ecological role of \textit{Gemmobacter} and \textit{Paucibacter}, which future research should seek to empirically test.}

{\color{blue}In addition to these observations about \textit{Gemmobacter} and \textit{Paucibacter}, compositional graphical lasso also finds insightful patterns about \textit{Photobacterium}, a genus which contains pathogens of marine fish \citep{romalde2002photobacterium, osorio2018photobacterium}, though its role as a pathogen in zebrafish is less clear.} However, as prior work has cultured \textit{Photobacterium} from the zebrafish gut \citep{cantas2012culturable} and other studies have demonstrated that \textit{Photobacterium} elicits toxicity to zebrafish cells in culture \citep{osorio2018photobacterium}, observations collectively point to \textit{Photobacterium} as a potential pathogen or pathobiont of zebrafish. Consistent with this speculation, prior work found that \textit{Photobacterium} statistically explained variation in \textit{P. tomentosa} worm burden as well as worm infection induced intestinal inflammation \citep{gaulke2019longitudinal}. We sought to determine if our interaction network analysis could clarify the role of \textit{Photobacterium} in the zebrafish gut microbiome, especially as it relates to parasite infection.

Interestingly, \textit{Photobacterium} is identified as an interaction hub (i.e., a relatively high degree node) in all networks assembled from all methods we evaluated. However, compositional graphical lasso uniquely reveals that the number of interactions linked to \textit{Photobacterium} increases in infected fish (19 edges) as compared to uninfected fish (16 edges), whereas the other approaches observe the opposite pattern (graphical lasso: 11 versus 17 edges; neighborhood selection: 3 versus 5 edges). These differences in the \textit{Photobacterium} subgraph across approaches also includes variation in the specific taxa that \textit{Photobacterium} is inferred to interact with. {\color{blue}To better understand changes in the set of \textit{Photobacterium} interactions, we extracted the first-degree sub-graph connected to \textit{Photobacterium} that each of the three methods produced for both the infected and uninfected host interaction networks. Compositional graphical lasso alone estimates eight interactions in the \textit{Photobacterium}-specific infected host network, and six of these eight genera have previously been linked to parasite exposure, disease burden, or histopathology score.} Two of these parasite etiology-linked taxa are \textit{Aeromonas} and \textit{Gemmobacter}, genera which are notable because they contain microbes that elicit pathogenic or pathobiotic phenotypes. For example, the \textit{Aeromonas} genus includes well-characterized and abundant opportunistic pathogens, such as \textit{A. hydrophila} \citep{saraceni2016establishment}, and the \textit{Gemmobacter} genus, as noted above, has been shown to opportunistically increase abundance \citep{huang2020exposure} in dysbiotic fish gut communities which display impeded overall gut function. {\color{blue}Given that \textit{Photobacterium} proliferates in the infected host gut, these observations suggest that \textit{Photobacterium} may work alongside and even promote the growth of other pathobiotic taxa to disrupt the composition of the gut microbiome and induce dysbiosis upon infection.}

Based on these collective observations, we hypothesize that \textit{Photobacterium} is an intestinal pathobiont of the zebrafish gut and contributes to dysbiosis in infected fish. \textit{Photobacterium} is a relatively prevalent genus in the zebrafish intestines, even in uninfected fish. However, \textit{P. tomentosa} infection links to an increase in the relative abundance of \textit{Photobacterium}, which also positively associates with intestinal hyperplasia \citep{gaulke2019longitudinal}, possibly due to the cytotoxic effect of \textit{Photobacterium}. Given that \textit{Photobacterium} is also an interaction hub whose influence on the community increases in infected fish, at least according to compositional graphic lasso, infection induced changes in \textit{Photobacterium} relative abundance may drive additional changes in the success of other taxa in the microbiome, including opportunistic pathogens and pathobionts like members of \textit{Aeromonas} and \textit{Gemmobacter}. Going forward, future studies should seek to experimentally validate our novel hypothesis about the pathobiotic role of \textit{Photobacterium} in the zebrafish gut, including its impact on the rest of the microbial community and its role in infection induced tissue damage. If our hypothesis is accurate, \textit{Photobacterium} may serve as an important model taxon for discerning how the gut microbiome and helminths interact to impact infection outcomes.

{\color{blue}Collectively, these results provide evidence for our hypothesis that in the case of intestinal helminth infection, dysbiosis may be defined by not only a change in the composition of the gut microbiome, but a restructuring in how key members of the microbial community interact with the rest of the microbiota.  Notably, these patterns were only observed by compositional graphical lasso, as the other methods did not observe unique evidence of such infection-associated inversions for \textit{Gemmobacter}, \textit{Paucibacter}, or \textit{Photobacterium}.

Finally, we visualize the six genus interaction networks, three for infected fish and three for uninfected fish (Figure \ref{fish_StARS}). For a better visualization, we only keep the top $100$ edges for the networks resulted from compositional graphical lasso and graphical lasso as they are much denser than the ones from neighborhood selection. Their edges are ranked in the exactly same way as in the \textit{Tara} Oceans Study, first by selection probability and then by edge weight (see Figure \ref{tara_StARS} in the supplementary materials for details). For all networks, darker blue implies higher magnitude in the absolute value of partial correlation.

\begin{figure}[htbp]
\centering
\includegraphics[scale=0.8]{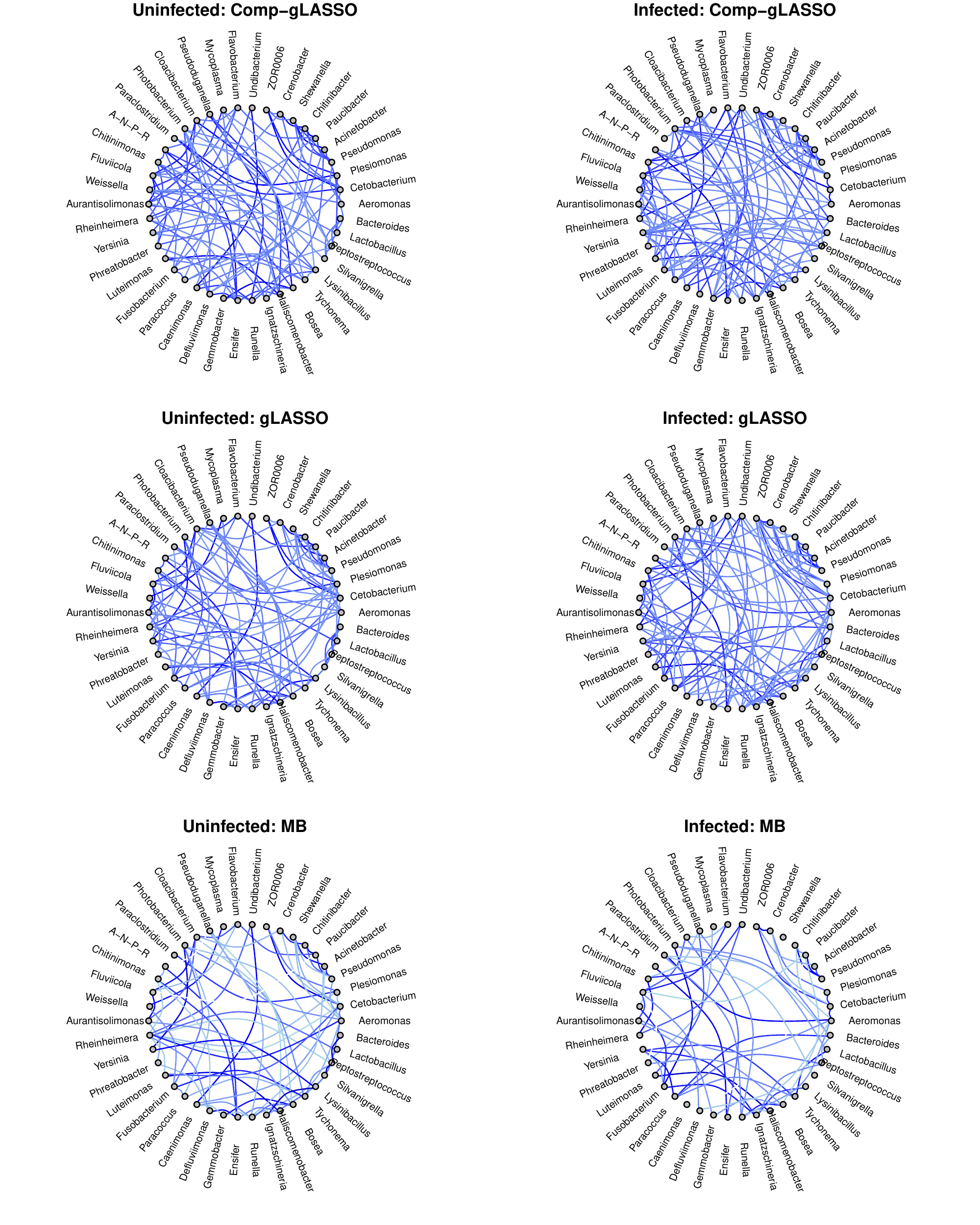}
\caption{Inferred networks from each method with edges filtered by selection probability and ranked by edge weight separately for the groups of uninfected and infected fish. In each network, darker blue implies stronger (larger in absolute value) edge weight.}
\label{fish_StARS}
\end{figure}}

\section{Discussion}
\label{Sec-5}

A growing body of work points to the gut microbiome as an agent to treat, diagnose, and prevent human diseases. The innovation of these utilities requires foundational knowledge about the pathobiotic {\color{blue}or probiotic} role that the gut microbiome play in the development and progression of the diseases. While most studies have focused on investigating how the abundance of gut microbes covary with a disease, such as infection \citep{gaulke2019longitudinal}, how the microbial interactions associate with the diseases is largely unknown. Understanding how the microbial interactions associate with a disease is critical in identifying microbes as a pathobiont {\color{blue}or probiotic}, which are candidates for potential drug or diagnostic test.

This work focuses on gaining a better understanding of the underlying role that microorganisms play in their communities by constructing microbial interaction networks. We propose a novel method called compositional graphical lasso that simultaneously accounts for the following key features of microbiome abundance data:  (a) the data are compositional counts and only carry information about relative abundances of the taxa; (b) the observed relative abundance is subject to heteroscedasticity as its variance depends on the varying sequence depth across samples; (c) the data are high-dimensional as the number of taxa are often larger than the number of samples. We have demonstrated the advantages of our approach over previously proposed methods in simulations and on a benchmark dataset.


We apply compositional graphical lasso to the data from the Zebrafish Parasite Infection Study, which used a parasite infection model to identify gut microbiota that positively and negatively associate with infection burden \citep{gaulke2019longitudinal}. {\color{blue}Our approach identified method-specific changes in interaction degree between infected and uninfected individuals for three taxa, \textit{Photobacterium}, \textit{Gemmobacter}, and \textit{Paucibacter}. Further investigation of these method-specific taxa interaction changes reveals their biological plausibility, and provides insight into their relevance in the context of parasite-linked changes in the zebrafish gut microbiome. In particular, we hypothesize that \textit{Photobacterium} and \textit{Gemmobacter} are pathobionts of zebrafish gut for parasitic infection and that \textit{Paucibacter} is a probiotic. Future studies should seek to experimentally validate their ecological roles in the zebrafish gut, including their impacts on the rest of the microbial community and their roles in infection induced tissue damage.}


It is noteworthy that compositional graphical lasso requires one to choose a reference taxon. As a general rule of thumb, we recommend choosing a ``common'' taxon that has a high average relative abundance and relatively few zero counts across samples, as its true relative abundance serves as the denominator in the additive log-ratio transformation. In practice, we also suggest that a user choose a taxon that is not of direct interest to investigate, as the reference taxon will not be represented in the resultant network. {\color{blue}It is also of interest to study how much the choice of the reference taxon may affect the estimated network, and whether some robustness could be guaranteed across different choices of the reference. Because the robustness against the choice of the reference is so important not just for compositional graphical lasso, but for many other network estimation methods as well, we have chosen to devote a major effort to this issue in a separate ongoing project. In this ongoing work, we have theoretically established the reference-invariance property of the inverse covariance matrix of a class of additive log-ratio transformation-based models, including the logistic normal multinomial model as a special case. However, as it is beyond the scope of this paper, this invariance property will be detailed in the future.}

\bibliographystyle{asa}
\bibliography{reference}

\newpage
\setcounter{page}{1} 

\setcounter{table}{0}
\renewcommand{\thetable}{S\arabic{table}}

\setcounter{figure}{0}
\renewcommand{\thefigure}{S\arabic{figure}}

\appendix

\section*{Supplementary Materials}

\subsection*{Proof of Theorem \ref{Thm-1}}

\begin{proof}
The conclusion in Theorem \ref{Thm-1} is directly implied by Theorem 4.1(c) in \cite{tseng2001convergence}, for which we need to verify a few regularity conditions as follows. 

First, $\ell_0(\bz_1,\ldots,\bz_n,\bmu,\bOmega)$ in (\ref{objective.function.inseparable}) is regular at each point in its domain. This is true because $\mathrm{dom}(\ell_0)$ is open and $\ell_0$ is differentiable and all its partial derivatives exists.

Second, the sublevel set $\{(\bz_1,\ldots,\bz_n,\bmu,\bOmega): \ell(\bz_1,\ldots,\bz_n,\bmu,\bOmega) \le \ell(\bz_1^{(0)},\ldots,\bz_n^{(0)},\bmu^{(0)},\bOmega^{(0)})\}$ is compact and that $\ell$ in (\ref{objective.function}) is continuous on this sublevel set. The continuity part is obvious and we just need to argue the compactness of the sublevel set. Since $\ell(\bz_1,\ldots,\bz_n,\bmu,\bOmega)$ is a continuous function, the compactness of this sublevel set is actually implied by the coerciveness of $\ell(\bz_1,\ldots,\bz_n,\bmu,\bOmega)$ \citep[Lemma 8.3]{calafiore2014optimization}, which we are going to argue as follows.

As argued previously, $\ell(\bz_1,\ldots,\bz_n,\bmu,\bOmega) = \ell_0(\bz_1,\ldots,\bz_n,\bmu,\bOmega) + \ell_1(\bz_1,\ldots,\bz_n) + \ell_2(\bOmega)$ as in (\ref{objective.function.inseparable})--(\ref{objective.function.separable.2}). On the one hand, let's argue that $\ell_0(\bz_1,\ldots,\bz_n,\bmu,\bOmega) + \ell_2(\bOmega)$ is bounded below as follows.
\begin{align*}
\ell_0(\bz_1,\ldots,\bz_n,\bmu,\bOmega) + \ell_2(\bOmega) 
& = -\frac12 \log[\det(\bOmega)] + \frac{1}{2n} \sum_{i=1}^n (\bz_i - \bmu)' \bOmega (\bz_i - \bmu) + \lambda \|\bOmega\|_1 \\
& \geq -\frac12 \log[\det(\hat{\bOmega})] + \frac{1}{2n} \sum_{i=1}^n (\bz_i - \bmu)' \hat{\bOmega} (\bz_i - \bmu) + \lambda \| \hat{\bOmega} \|_1 \\
& \geq -\frac12 \log[\det(\hat{\bOmega})]\\
& \geq - \frac 12 K \log \frac{K}{\lambda},
\end{align*}
where 
\[\hat{\bOmega} = \argmin_{\bOmega} -\frac12 \log[\det(\hat{\bOmega})] + \frac{1}{2n} \sum_{i=1}^n (\bz_i - \bmu)' \hat{\bOmega} (\bz_i - \bmu) + \lambda \| \hat{\bOmega} \|_1,\] 
and the last inequality follows because $\hat{\bOmega}$ is unique and has positive eigenvalues bounded above by $\frac{K}{\lambda}$ \citep{banerjee2008model}.

On the other hand, we argue that $\ell_1(\bz_1,\ldots,\bz_n)$ is a coercive function of $\bz_1,\ldots,\bz_n$. This is because $\ell_1(\bz_1,\ldots,\bz_n)$ is proper, closed (as implied by continuity), strictly convex and has a unique minimizer (with a positive definite Hessian matrix, see Section \ref{Sec-2.3}), thus every sublevel set of $\ell_1$ is bounded, following Corollary 8.7.1 from \citet{rockafellar1970convex}. In addition, every sublevel set of $\ell_1$ is closed due to the fact that $\ell_1$ is continuous \citep[Theorem 7.1]{rockafellar1970convex}. Therefore, every sublevel set of $\ell_1$ is compact, implying the coerciveness of $\ell_1$ \citep[Lemma 8.3]{calafiore2014optimization}.

Combining that $\ell_0(\bz_1,\ldots,\bz_n,\bmu,\bOmega) + \ell_2(\bOmega)$ is bounded below and that $\ell_1(\bz_1,\ldots,\bz_n)$ is a coercive function of $\bz_1,\ldots,\bz_n$, then if $\|(\bz_1',\ldots,\bz_n')\| \to \infty$, $\ell(\bz_1,\ldots,\bz_n,\bmu,\bOmega) \to \infty$.
Thus, to show that $\ell(\bz_1,\ldots,\bz_n,\bmu,\bOmega)$ is a coercive function, we just need to show that $g(\bmu, \bOmega) = \ell_0(\bz_1,\ldots,\bz_n,\bmu,\bOmega) + \ell_2(\bOmega)$ is a coercive function of $\bmu$ and $\bOmega$ for fixed and finite $\bz_1,\ldots,\bz_n$.

Note that for fixed and finite $\bz_1,\ldots,\bz_n$, 
\begin{align*}
g(\bmu, \bOmega)
& = -\frac12 \log[\det(\bOmega)] + \frac{1}{2n} \sum_{i=1}^n (\bz_i - \bmu)' \bOmega (\bz_i - \bmu) + \lambda \|\bOmega\|_1 \\
& \ge -\frac12 \sum_{j=1}^K  \log[\kappa_j(\bOmega)] + \frac{1}{2n} \kappa_{\min}(\bOmega) \sum_{i=1}^n \|\bz_i - \bmu\|^2 + \lambda \kappa_{\max}(\bOmega) \\
& \ge \max(I_1, I_2),
\end{align*}
where $\kappa_j(\bOmega)$, $\kappa_{\min}(\bOmega)$, and $\kappa_{\max}(\bOmega)$ denote the $j$th, minimum, and maximum eigenvalues of $\bOmega$, respectively, and 
\begin{align*}
I_1 &= \frac{1}{2n} \kappa_{\min}(\bOmega) \sum_{i=1}^n \|\bz_i - \bmu\|^2 + \lambda \kappa_{\max}(\bOmega) -\frac12 K  \log[\kappa_{\max}(\bOmega)], \\
I_2 &= -\frac12 \log[\kappa_{\min}(\bOmega)] + \lambda \kappa_{\max}(\bOmega) - \frac12 (K-1)\log[\kappa_{\max}(\bOmega)].
\end{align*}

For $L = K$ or $L = K - 1$, $\lambda\kappa_{\max}(\bOmega) - \frac12 L \log[\kappa_{\max}(\bOmega)] \ge 0$ when $\kappa_{\max}(\bOmega) \to \infty$ and $\lambda\kappa_{\max}(\bOmega) - \frac12 L \log[\kappa_{\max}(\bOmega)]$ is bounded below when $\kappa_{\max}(\bOmega)$ is bounded above. Therefore, if $\kappa_{\min}(\bOmega)$ is bounded away from 0, then $I_1 \to \infty$ when $\|\bmu\| \to \infty$ because $\frac{1}{2n} \kappa_{\min}(\bOmega) \sum_{i=1}^n \|\bz_i - \bmu\|^2 \to \infty$; if $\kappa_{\min}(\bOmega) \to 0$, then $I_2 \to \infty$ because $-\frac12 \log[\kappa_{\min}(\bOmega)] \to \infty$. In summary, for fixed and finite $\bz_1,\ldots,\bz_n$, we have $g(\bmu, \bOmega) \to \infty$ when $\|\bmu\| \to \infty$ regardless of $\bOmega$.

Thus, to show that $g(\bmu, \bOmega)$ is a coercive function of $\bmu$ and $\bOmega$ for fixed and finite $\bz_1,\ldots,\bz_n$, it suffices to show that $g(\Omega) = \ell_0(\bz_1,\ldots,\bz_n,\bmu,\bOmega) + \ell_2(\bOmega)$ is a coercive function of $\bOmega$ for fixed and finite $\bz_1,\ldots,\bz_n$ and $\mu$. As $g(\Omega)$ is the graphical lasso objective function, it is proper, closed, convex, and has a unique minimizer \citep{banerjee2008model, friedman2008sparse}, thus every sublevel set of $g$ is bounded, again following Corollary 8.7.1 from \citet{rockafellar1970convex}. In addition, every sublevel set of $g$ is closed due to the continuity of $g$. Therefore, every sublevel set of $g$ is compact, implying the coerciveness of $g$ \citep[Lemma 8.3]{calafiore2014optimization}. This concludes the proof of the coerciveness of the function $\ell(\bz_1,\ldots,\bz_n,\bmu,\bOmega)$.

Third, $\ell(\bz_1,\ldots,\bz_n,\bmu,\bOmega)$ has at most one minimum in its second block of parameters, i.e., $\bmu$. This is because the Hessian matrix for $\bmu$ is $\bOmega$ which is positive definite. 

The conclusion of Theorem \ref{Thm-1} is thus proved as we have verified all regularity conditions in Theorem 4.1(c) in \cite{tseng2001convergence}.
\end{proof}

\newpage 

\begin{figure}[htbp]
\centering
\includegraphics[width=\textwidth]{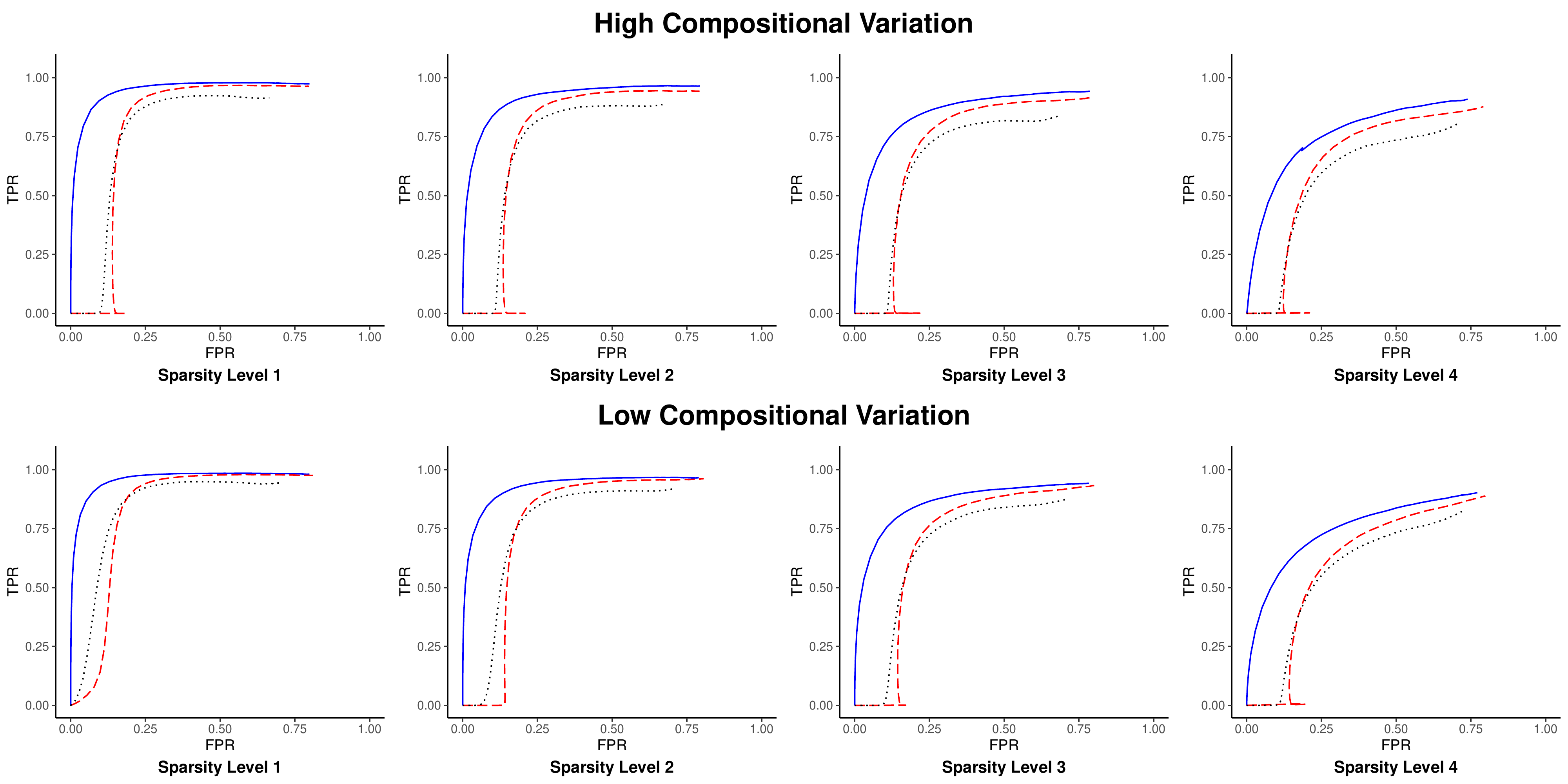}
\caption{ROC curves for compositional graphical lasso (Comp-gLASSO), graphical lasso (gLASSO) and neighborhood selection (MB). Solid blue: Comp-gLASSO; dashed red: gLASSO; dotted black: MB. Sparsity levels 1--4: from the least sparse to the most sparse.}
\label{ROC_MB_Sparse}
\end{figure}

\begin{figure}[htbp]
\centering
\includegraphics[width=\textwidth]{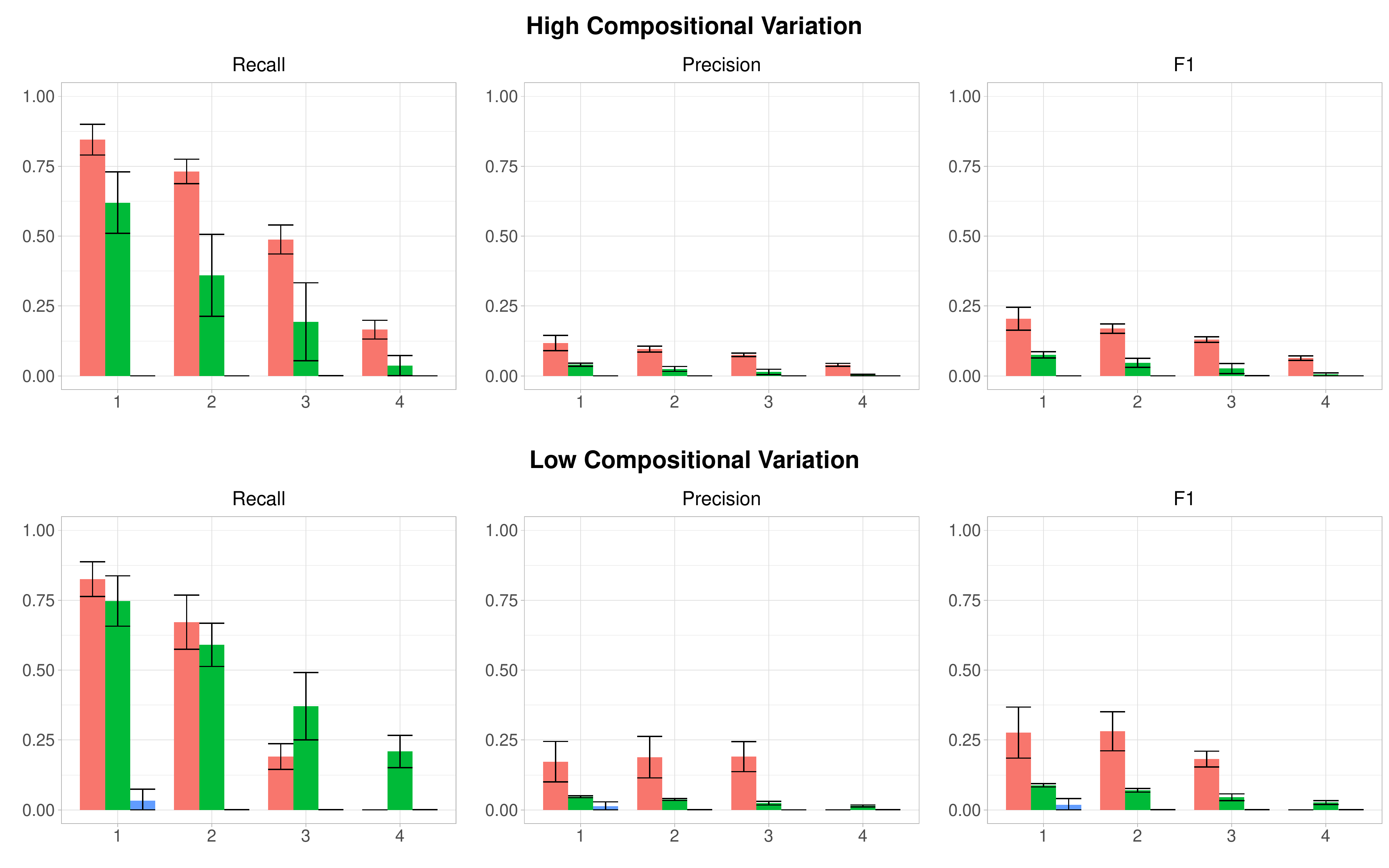}
\caption{Recall, precision and F1 score for the network selected by StARS for compositional graphical lasso (Comp-gLASSO), graphical lasso (gLASSO) and neighborhood selection (MB). Red (left): Comp-gLASSO; green (middle): gLASSO; blue (right): MB. 1--4: sparsity levels 1--4 from the least sparse to the most sparse.}
\label{Recall.Precision.F1.Sparse}
\end{figure}

\subsection*{Simulations: Figures \ref{ROC_MB_Sparse} and \ref{Recall.Precision.F1.Sparse}}

\newpage 

\begin{figure}[htbp]
\centering
\includegraphics[scale=0.8]{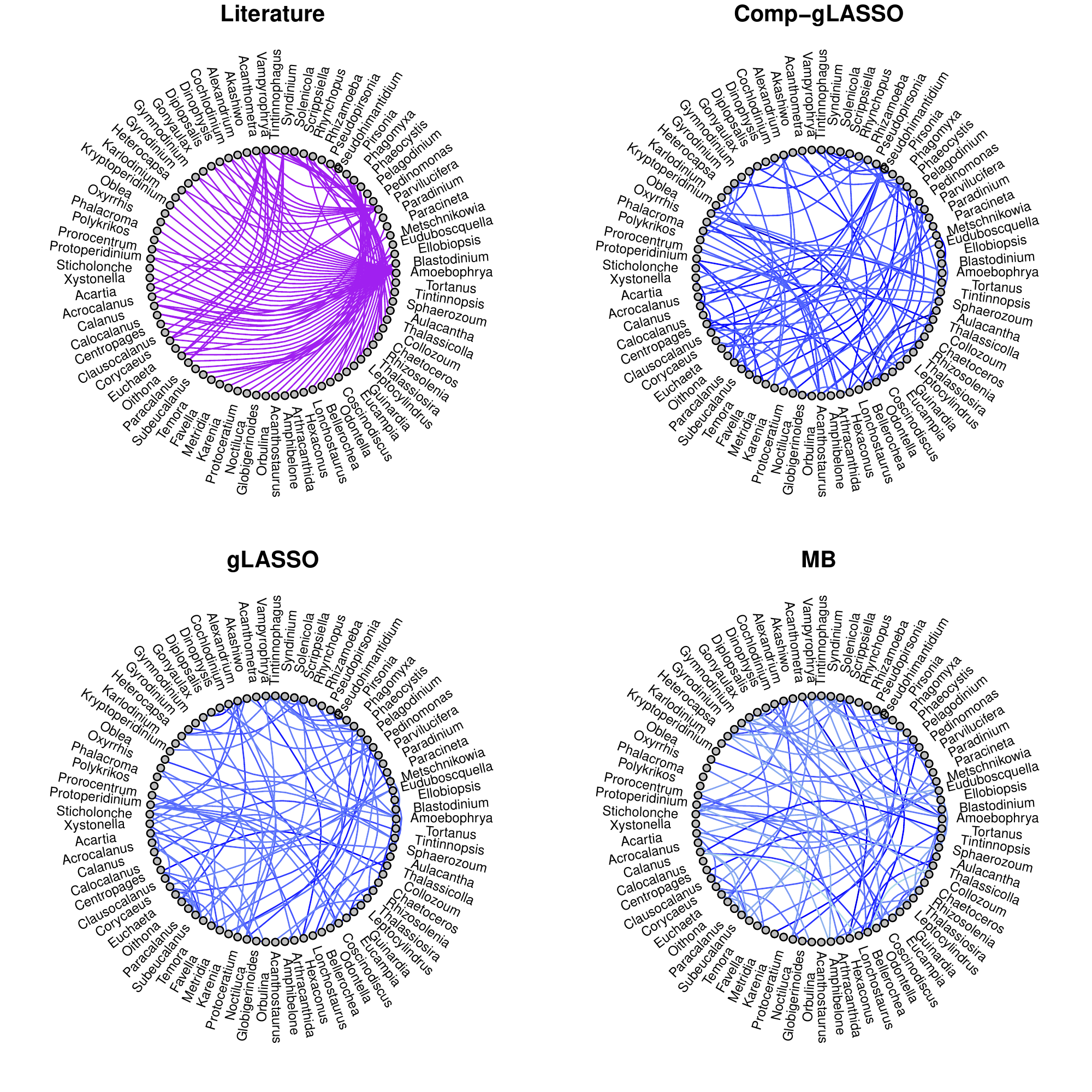}
\caption{Inferred networks from each method with edges filtered by selection probability and ranked by edge weight, in comparison with the 91 interactions reported in literature. In each network, darker blue implies stronger (larger in absolute value) edge weight.}
\label{tara_StARS}
\end{figure}

\subsection*{\textit{Tara} Oceans Project: Figure \ref{tara_StARS} and Table \ref{common_genus_pairs}}

We visualize the three genus interaction networks (from compositional graphical lasso, graphical lasso, and neighborhood selection) in company with the network of the literature validated interactions. For a better visualization, we only keep the top $100$ edges that are ranked by the following two criteria: (a) selection probability, the proportion of times that an edge is selected from the $N$ subsamples in StARS and (b) edge weight, the absolute value of the partial correlation that is defined as $|\hat\omega_{ij}|/\sqrt{\hat\omega_{ii}\hat\omega_{jj}}$ where $\hat\omega_{ij}$ is the $(i,j)$ entry of the estimated inverse covariance matrix $\hat\Omega$. Specifically, the edges are first ranked by selection probability, and the edges with the same selection probabilities are further ranked by edge weight. For the networks from all three methods, darker blue implies higher magnitude in the absolute value of partial correlation. 

We can see from Figure \ref{tara_StARS} that, though still different, the networks estimated by the three algorithms have apparent similarity in the predicted edges and their edge weights, e.g., the genus pairs ``\textit{Centropages} - \textit{Thalassicolla}" and ``\textit{Acanthometra} - \textit{Hexaconus}" are in dark blue for all three methods. On the other hand, there are very few overlaps between those top 100 edges and the known interactions from literature. Since our current knowledge of the genus-level interactions are still limited, the edges that have not been reported from literature but enjoys higher selection probability and larger weight might suggest promising new eukaryotic interactions that deserve biological validations. There are actually 39 common edges from the top 100 edges from the three estimated networks. We provide the list of these common genus-genus interactions (Table \ref{common_genus_pairs}) for the interested readers.

\begin{table}[H]
\centering
\begin{tabular}{rll}
  \hline
 & Genus 1 & Genus 2\\ 
  \hline
1     &   Gonyaulax  &     Alexandrium\\
2  & Thalassicolla    &   Centropages\\
3   &   Sphaerozoum    &     Collozoum\\
4   &     Hexaconus   &   Acanthometra\\
5   &   Pedinomonas    &       Karenia\\
6   &  Lonchostaurus    &    Aulacantha\\
7   &      Eucampia  &   Coscinodiscus\\
8    &   Phagomyxa     &    Noctiluca\\
9    &       Temora    &   Centropages\\
10    &       Oblea &    Coscinodiscus\\
11      &   Oithona   &    Alexandrium\\
12     &   Orbulina &  Globigerinoides\\
13   & Vampyrophrya    &     Syndinium\\
14   &   Paracineta   &       Euchaeta\\
15      &  Tortanus  &         Calanus\\
16   & Scrippsiella     &   Gyrodinium\\
17 & Pseudopirsonia       &  Pirsonia\\
18    & Heterocapsa     &     Akashiwo\\
19   &Thalassicolla   &     Collozoum\\
20    &  Rhynchopus   &    Phaeocystis\\
21  &Tintinnophagus   &      Pirsonia\\
22   &Thalassiosira    &  Rhizosolenia\\
23  & Prorocentrum   &      Hexaconus\\
24  & Pseudopirsonia   &     Paradinium\\
25  & Metschnikowia    &       Acartia\\
26  & Leptocylindrus      &    Eucampia\\
27   &    Odontella   &    Chaetoceros\\
28 & Globigerinoides   &      Corycaeus\\
29  & Tintinnophagus   &    Rhynchopus\\
30   & Metschnikowia   &    Aulacantha\\
31  &  Pelagodinium    &       Favella\\
32  &  Prorocentrum   & Euduboscquella\\
33   & Tintinnopsis   &  Thalassiosira\\
34     &  Collozoum    &       Acartia\\
35  &  Subeucalanus  &     Sphaerozoum\\
36   & Vampyrophrya  & Pseudohimantidium\\
37   &     Oxyrrhis &   Arthracanthida\\
38     &     Temora    &   Acrocalanus\\
39  & Tintinnophagus    &   Amoebophrya\\
   \hline
\end{tabular}
\caption{\label{tab:genus pair in common}Genus-genus interactions that are commonly identified by compositional graphical lasso, graphical lasso, and neighborhood selection.}
\label{common_genus_pairs}
\end{table}

\newpage 

\subsection*{Zebrafish Parasite Infection Study: Table \ref{fish_hubs}}

\begin{sidewaystable}
\scriptsize
\begin{tabular}{llllllll}
\hline
Uninfected & Infected && Uninfected & Infected && Uninfected & Infected \\
\multicolumn{2}{c}{Comp-gLASSO} && \multicolumn{2}{c}{gLASSO} && \multicolumn{2}{c}{MB}\\
\cline{1-2} \cline{4-5}  \cline{7-8}
Plesiomonas (21)       & Photobacterium (19)    && Silvanigrella (21)     & Paucibacter (21)       && Rheinheimera (9)      & Fluviicola (7)        \\
Paraclostridium (18)   & Weissella (18)         && Peptostreptococcus (21)& Phreatobacter (18)     && Aeromonas (8)         & Aeromonas (6)         \\
Aeromonas (17)         & Gemmobacter (18)       && Aeromonas (20)         & Fusobacterium (17)     && Cetobacterium (8)     & Paucibacter (5)       \\
Paucibacter (17)       & Fusobacterium (17)     && Plesiomonas (20)       & Cetobacterium (16)     && Runella (6)           & Runella (5)           \\
Photobacterium (16)    & Paucibacter (15)       && Paraclostridium (19)   & Cloacibacterium (16)   && Peptostreptococcus (6)& Cetobacterium (4)     \\
Fusobacterium (16)     & Yersinia (15)          && Pseudoduganella (18)   & Weissella (16)         && Pseudomonas (5)       & Paraclostridium (4)   \\
Lactobacillus (16)     & Crenobacter (14)       && Rheinheimera (18)      & ZOR0006 (15)           && Cloacibacterium (5)   & Yersinia (4)          \\
Rheinheimera (15)      & Aurantisolimonas (14)  && Bacteroides (18)       & Flavobacterium (15)    && Photobacterium (5)    & Fusobacterium (4)     \\
Yersinia (15)          & Paracoccus (14)        && Undibacterium (17)     & Yersinia (15)          && Haliscomenobacter (5) & Paracoccus (4)        \\
Silvanigrella (15)     & Ignatzschineria (14)   && Photobacterium (17)    & Shewanella (14)        && Tychonema (5)         & Ignatzschineria (4)   \\
Bacteroides (15)       & Lysinibacillus (14)    && Runella (17)           & Aurantisolimonas (14)  && Plesiomonas (4)       & Bosea (4)             \\
Chitinibacter (14)     & Plesiomonas (13)       && Paucibacter (16)       & Aeromonas (13)         && Acinetobacter (4)     & Bacteroides (4)       \\
Weissella (14)         & Acinetobacter (13)     && Chitinimonas (16)      & Acinetobacter (13)     && Paucibacter (4)       & Pseudomonas (3)       \\
Paracoccus (14)        & Chitinibacter (13)     && Yersinia (16)          & Crenobacter (13)       && Shewanella (4)        & Chitinibacter (3)     \\
Peptostreptococcus (14)& Rheinheimera (13)      && Paracoccus (16)        & A-N-P-R (13)           && Pseudoduganella (4)   & Crenobacter (3)       \\
Pseudomonas (13)       & Cetobacterium (12)     && Caenimonas (16)        & Pseudoduganella (12)   && Chitinimonas (4)      & ZOR0006 (3)           \\
Acinetobacter (13)     & Cloacibacterium (12)   && Lactobacillus (16)     & Caenimonas (12)        && Fluviicola (4)        & Pseudoduganella (3)   \\
Pseudoduganella (13)   & Phreatobacter (12)     && Luteimonas (15)        & Gemmobacter (12)       && Aurantisolimonas (4)  & Cloacibacterium (3)   \\
Cloacibacterium (13)   & Tychonema (12)         && Ignatzschineria (15)   & Ignatzschineria (12)   && Fusobacterium (4)     & Photobacterium (3)    \\
Chitinimonas (13)      & Bacteroides (12)       && Lysinibacillus (15)    & Lysinibacillus (12)    && Caenimonas (4)        & Chitinimonas (3)      \\
Ignatzschineria (13)   & Aeromonas (11)         && Cetobacterium (14)     & Lactobacillus (12)     && Defluviimonas (4)     & Aurantisolimonas (3)  \\
Lysinibacillus (13)    & Shewanella (11)        && Chitinibacter (14)     & Pseudomonas (11)       && Ensifer (4)           & Phreatobacter (3)     \\
Shewanella (12)        & Undibacterium (11)     && Shewanella (14)        & Chitinibacter (11)     && Bosea (4)             & Defluviimonas (3)     \\
ZOR0006 (12)           & Chitinimonas (11)      && Cloacibacterium (14)   & Mycoplasma (11)        && Chitinibacter (3)     & Tychonema (3)         \\
Flavobacterium (12)    & Lactobacillus (11)     && Phreatobacter (14)     & Photobacterium (11)    && Crenobacter (3)       & Peptostreptococcus (3)\\
Mycoplasma (12)        & ZOR0006 (10)           && Gemmobacter (14)       & Fluviicola (11)        && Flavobacterium (3)    & Plesiomonas (2)       \\
Tychonema (12)         & Pseudoduganella (10)   && Bosea (14)             & Rheinheimera (11)      && A-N-P-R (3)           & Acinetobacter (2)     \\
Undibacterium (11)     & A-N-P-R (10)           && Tychonema (14)         & Paracoccus (11)        && Luteimonas (3)        & Undibacterium (2)     \\
Crenobacter (10)       & Mycoplasma (9)         && Pseudomonas (13)       & Runella (11)           && Paracoccus (3)        & Flavobacterium (2)    \\
Fluviicola (10)        & Fluviicola (9)         && Acinetobacter (13)     & Haliscomenobacter (11) && Gemmobacter (3)       & Weissella (2)         \\
Defluviimonas (10)     & Luteimonas (9)         && A-N-P-R (13)           & Bosea (11)             && Lysinibacillus (3)    & Rheinheimera (2)      \\
Ensifer (10)           & Defluviimonas (9)      && Fluviicola (13)        & Plesiomonas (10)       && ZOR0006 (2)           & Luteimonas (2)        \\
Cetobacterium (9)      & Ensifer (9)            && Weissella (13)         & Undibacterium (10)     && Undibacterium (2)     & Gemmobacter (2)       \\
A-N-P-R (9)            & Peptostreptococcus (9) && ZOR0006 (12)           & Paraclostridium (10)   && Weissella (2)         & Ensifer (2)           \\
Aurantisolimonas (9)   & Pseudomonas (8)        && Flavobacterium (12)    & Chitinimonas (10)      && Yersinia (2)          & Haliscomenobacter (2) \\
Luteimonas (9)         & Flavobacterium (8)     && Mycoplasma (12)        & Defluviimonas (10)     && Phreatobacter (2)     & Lactobacillus (2)     \\
Runella (9)            & Caenimonas (8)         && Defluviimonas (12)     & Ensifer (10)           && Ignatzschineria (2)   & Mycoplasma (1)        \\
Haliscomenobacter (9)  & Haliscomenobacter (8)  && Haliscomenobacter (12) & Tychonema (10)         && Lactobacillus (2)     & A-N-P-R (1)           \\
Caenimonas (8)         & Bosea (8)              && Crenobacter (11)       & Silvanigrella (10)     && Mycoplasma (1)        & Lysinibacillus (1)    \\
Gemmobacter (8)        & Silvanigrella (8)      && Aurantisolimonas (10)  & Luteimonas (9)         && Paraclostridium (1)   & Silvanigrella (1)     \\
Bosea (8)              & Paraclostridium (6)    && Fusobacterium (10)     & Peptostreptococcus (9) && Silvanigrella (1)     & Shewanella (0)        \\
Phreatobacter (7)      & Runella (5)            && Ensifer (9)            & Bacteroides (9)        && Bacteroides (1)       & Caenimonas (0)        \\
\hline
\end{tabular}
\caption{The ranks of genera in each degree distribution (in descending order) for uninfected fish and infected fish separately, from compositional graphical lasso (Comp-gLASSO), graphical lasso (gLASSO), and neighborhood selection (MB). The numbers in the parentheses are the corresponding degrees of the genera.}
\label{fish_hubs}
\end{sidewaystable}




\end{document}